\documentclass[10pt,twocolumn,oneside,a4paper,final]{IEEEtran}

\ifCLASSINFOpdf
\else
\fi
\makeatletter
\def\ps@headings{%
\def\@oddhead{\mbox{}\scriptsize\rightmark \hfil \thepage}%
\def\@evenhead{\scriptsize\thepage \hfil \leftmark\mbox{}}%
\def\@oddfoot{}%
\def\@evenfoot{}}
\makeatother 
\IEEEoverridecommandlockouts
\def\BibTeX{{\rm B\kern-.05em{\sc i\kern-.025em b}\kern-.08em
    T\kern-.1667em\lower.7ex\hbox{E}\kern-.125emX}}

\usepackage[utf8]{inputenc}
\usepackage[acronym]{glossaries}
\glsdisablehyper
\newacronym{urllc}{URLLC}{Ultra reliable low latency communication}
\newacronym{hrllc}{HRLLC}{hyper reliable low latency communication}
\newacronym{rl}{RL}{reinforcement learning}
\newacronym{drl}{DRL}{deep reinforcement learning}
\newacronym{harq}{HARQ}{hybrid automatic repeat request}
\newacronym{ppo}{PPO}{proximal policy optimization}
\newacronym{ccharq}{CC-HARQ}{chase combining HARQ}
\newacronym{irharq}{IR-HARQ}{incremental redundancy HARQ}
\newacronym{csi}{CSI}{channel state information}
\newacronym{snr}{SNR}{signal-to-noise ratio}
\newacronym{qos}{QoS}{quality of service}
\newacronym{3gpp}{3GPP}{3rd Generation Partnership Project}
\newacronym{mrc}{MRC}{maximal-ratio-combining}
\newacronym{cdf}{CDF}{cumulative distribution function}
\newacronym{tx}{TX}{transmitter}
\newacronym{rx}{RX}{receiver}
\newacronym{podd}{PODD}{proactive outdated data dropping}
\newacronym{dvp}{DVP}{delay violation probability}
\newacronym{noma}{NOMA}{non-orthogonal multiple access}
\newacronym{bs}{BS}{base station}
\newacronym{ue}{UE}{user equipment}
\newacronym{mse}{MSE}{mean square error}
\newacronym{ckm}{CKM}{channel knowledge map}
\newacronym{maml}{MAML}{model agnostic meta learning}
\newacronym{gp}{GP}{Gaussian process}
\newacronym{iot}{IoT}{Internet of Things}
\newacronym{mdp}{MDP}{Markov decision process}

\newcommand{\mbf}[1]{\mathbf{#1}}
\newcommand{\mbs}[1]{\boldsymbol{#1}}

\usepackage{mathrsfs}
\usepackage{cite}
\usepackage{amsmath,amssymb,amsfonts,amsthm}
\usepackage{graphicx}
\usepackage{textcomp}
\usepackage[dvipsnames]{xcolor}
\usepackage{algorithmic}
\usepackage{algorithm}
\usepackage{epstopdf}
\usepackage{subfigure}
\usepackage{caption}
\usepackage{stfloats}
\usepackage{hyperref}
\usepackage{float}
\usepackage{graphicx} % Required for inserting images

\title{{{Fast Transmission Control Adaptation for URLLC via Channel Knowledge Map and Meta-Learning}}
\thanks{H. Peng and M. Tao are with the Department of Electronic Information and Electrical Engineering,  and the Cooperative Medianet Innovation Center (CMIC), Shanghai Jiao Tong University, Shanghai, China. (Emails: \{hs-peng, mxtao\}@sjtu.edu.cn) 

T. Kallehauge and P. Popovski are with the Department of  Electronic Systems, Aalborg University, Aalborg, Denmark. (Emails:  \{tkal, petarp\}@es.aau.dk)}
}
\author{Hongsen  Peng, Tobias Kallehauge, Meixia Tao,~\textit{Fellow, IEEE}, and Petar Popovski,~\textit{Fellow, IEEE}}

\begin{document}

\maketitle

\begin{abstract}
This paper considers methods for delivering ultra reliable low latency communication (URLLC) {to enable mission-critical  Internet of Things (IoT) services in} wireless environments with unknown channel distribution. The methods rely upon the historical channel gain samples of a few locations in a target area. We formulate a non-trivial transmission control adaptation problem across the target area under the URLLC constraints. {Then we propose two solutions to solve this problem. }
The \emph{first} is a power scaling scheme {in conjunction with the deep reinforcement learning (DRL) algorithm with the help of the channel knowledge map (CKM) without retraining}, where the CKM employs the spatial correlation of the channel characteristics from the historical channel gain samples. 
The \emph{second solution} is model agnostic meta-learning (MAML) based meta-reinforcement learning algorithm that is trained from the known channel gain samples following distinct channel distributions and can quickly adapt to the new environment within a few steps of gradient update. Simulation results indicate that the DRL-based algorithm can effectively meet the reliability requirement of URLLC under various quality-of-service (QoS) constraints. Then the adaptation capabilities of the power scaling scheme and meta-reinforcement learning algorithm are also validated. 
\end{abstract}

\section{Introduction}
\gls{urllc} is one of the main usage scenarios of 5G wireless networks and will be enhanced further to \gls{hrllc} in 6G networks \cite{8705373,lopes, 6GHRLLC,IMT2030}. %URLLC has both a very stringent end-to-end latency constraint and an ultra high reliability requirement \cite{8705373,lopes}. 
\gls{urllc} is envisioned to meet the extremely stringent \gls{qos} requirements so as to enable many mission-critical {\gls{iot}}  applications, such as autonomous driving, industrial automation, unmanned aerial vehicles (UAVs) control, smart grid, etc \cite{Bennis}. 
The latency of \gls{urllc} can be generally measured by the end-to-end latency at the link layer \cite{IMT2030, 8705373} (e.g. 1 ms), {which takes into account both transmission delay in the physical layer and the queueing delay at the transmitter buffer in the link layer. 
{Reliability is defined as the probability (e.g. 99.999\%) that the transmitter successfully delivers a finite-size data packet to the receiver within a targeted time interval}. The complement of this reliability is known as \gls{dvp}.
This cross-layer end-to-end approach considering all the potential delay and error sources is inevitably required for providing strong reliability and latency guarantees\cite{she,lopes}. 
The contradicting requirements of low latency and high reliability will bring in significant challenges in both the physical layer and the link layer of the wireless communication network\cite{Bennis}.

{At }the physical layer the challenge mainly lies in the transmission outage due to the stochastic channel fading. % achieving ultra/hyper reliable and low latency communication is very challenging. 
Adaptive transmission control by exploiting instantaneous \gls{csi} at the transmitter is an effective approach to overcome the transmission outage and improve reliability. However, acquiring instantaneous \gls{csi} at the transmitter is difficult.  On the one hand, channel estimation will bring in non-negligible latency due to pilot symbol transmission and estimated \gls{csi} feedback, which increases the overall latency and may be unacceptable for mission-critical~\gls{urllc}. On the other hand, the CSI at the transmitter may not be perfect due to channel estimation error or feedback error and may also be outdated due to the rapid channel variation. Such imperfect and outdated CSI at the transmitter can significantly degrade the effectiveness of adaptive transmission control\cite{icsi}. 
% On the oter hand,  Even though the estimated \gls{csi} is fed back to the transmitter side, errors may occur in this step. In the meanwhile, assuming the instantaneous channel coefficient stays exactly the same in the training phase and data transmission phase depends on the actual channel dynamic \cite{icsi}. 
Rather than relying on instantaneous \gls{csi} acquisition, the transmission protocol can utilize statistical channel knowledge to ensure reliability and latency in a probabilistic manner \cite{8673783}. In the context of ultra-reliable communication, special attention must be paid to modeling the \emph{tail distribution} of the channel in order to capture rare events such as deep fades. Traditional parametric channel models such as Rayleigh or Rician fading may lead to severe model mismatch in the tail distribution \cite{8673783}, so recent works have proposed leveraging non-parametric models to avoid this problem \cite{srm}.  

From the link layer perspective, transmission needs to take into account queueing information. There are works that analyze \gls{dvp} and delay-reliability tradeoff for \gls{urllc} wireless communication systems using various tools, such as effective capacity\cite{QiaoAug.2019,10495827}, stochastic network calculus\cite{schi}, extreme value theory\cite{EVT} and so on. These works usually call for some specific approximations to derive corresponding tractable upper bounds. Although these bounds can provide transmission control guidance, the tightness of the bounds cannot be strictly guaranteed, resulting in low power efficiency. In addition, usually these tools require not only {statistical channel knowledge} as prior, but also the instantaneous \gls{csi}. %, which is not applicable in scenarios where the channel statistics are not known.
% Recently, {\gls{drl}} approaches have been proven to be effective for solving the transmission control problem for \gls{urllc}\cite{She2021}. 
% In \cite{ Kasgari2021},  the authors proposed a \gls{drl} algorithm to solve the resource allocation problem with guaranteed reliability and latency. }  
% The work \cite{zhangli} proposed a \gls{drl}-based resource allocation algorithm for channel estimation and data transmission in multiple-input and single-output (MISO) ultra-reliable low-latency communication (URLLC) systems.  
% Therein, only the latency at the physical layer is considered without taking into account the queuing delay at the link layer in \cite{Kasgari2021,zhangli}.  
In \cite{Harq}, we have studied a power efficient \gls{harq}-enabled mechanism for \gls{urllc} systems. We proposed a \gls{drl} algorithm that optimizes the instantaneous coding rate and transmit power based on the queueing information and provides a tight assurance on \gls{dvp}. Therein, only the statistical \gls{csi} of the environment is needed at the transmitter side.  

Although relying on statistical channel knowledge can be beneficial for long-term reliability guarantees, estimating the statistics can be resource-intensive and only applies to stationary channel environment --- once a user moves, the channel statistics may change and the parameters need to be re-estimated.  
A promising direction to acquire channel statistics with significantly fewer resources is spatial prediction. It leverages historical transmissions in a network to predict the \gls{csi} distribution directly from a user's location. Spatial prediction assumes that the long-term statistics vary smoothly in space due to shared dominant paths, scatterers, etc in the propagation environment. Thus, channel samples collected from other users in the network can, when combined with location information, be expected to predict the channel statistics at a new, yet nearby, location. 
% \textcolor{blue}{Such assumptions are also required in radio maps \cite{ramap1,ramap2} relevant works, i.e., coverage maps, which typically model the average received \gls{snr} or received signal strength.}
Spatial prediction of channel statistics such as radio map \cite{ramap1,ramap2} is gaining increasing attention with the rise of environment-aware wireless communication for 6G networks \cite{Lima2021Sensing} and we see a new generation of radio maps with advanced capabilities in terms of the \gls{ckm} \cite{ckm,ckmtutorial}. 
Generally speaking, \gls{ckm} can be regarded as a generalization of traditional radio maps, i.e., coverage maps, which typically model the average received \gls{snr} or received signal strength in a target coverage area. 
%This \gls{ckm} goes beyond this paradigm through a site-specific database with the positions of transmitters and/or receivers. 
This \gls{ckm} also contains channel-related key information, e.g., delays/Dopplers/AoAs/AoDs of multi-paths or the complete channel impulse response. This information serves to improve environmental awareness and streamline or possibly eliminate the need for intricate real-time acquisition of \gls{csi}\cite{ckm}. 
% However, previous works utilizing \gls{ckm} capturing the average channel information cannot be applied in the \gls{urllc} systems directly due to the sensitivity of the decoding error probability of the transmission. 
Regarding the channel information beyond average statistics in \gls{ckm}, the work \cite{srm} introduced a novel non-parametric \gls{ckm} that predicts $\varepsilon$-quantiles of the channel gain through the Gaussian process and proposed a rate selection algorithm for given target reliability. 
Note that the above \gls{ckm} work \cite{srm} focuses on physical-layer adaptation only. 

This work aims to exploit CKM from the cross-layer design perspective to enhance the transmission control adaptability {for \gls{urllc} in mission-critical \gls{iot} systems without knowing channel distribution as prior.} 
Motivated by the \gls{ckm} works investigating the information from previously collected channel gain samples of different locations through spatial correlation, we also consider adopting meta-reinforcement learning to exploit this information and overcome the policy adaptation problem across the target area. % by extracting an efficient network initialization from a variety of tasks and then fast adapt to new locations. 
In summary, we aim to investigate how to adapt the transmission control policy for the devices within the target area with different channel statistics while following spatial correlations. We first utilize a power scaling scheme in conjunction with \gls{drl} algorithm with the aid of the \gls{ckm} constructed from the historically observed location-restricted channel gain samples. Then we employ a meta-reinforcement learning algorithm that learns an appropriate network parameter initialized from the observed channel gain samples following distinct channel distributions. This initialization can quickly adapt to the new environment within just a few gradient update steps. 
The main contributions of this work can be summarized as follows: 
\begin{itemize}
\item {%We consider a test area where a limited number of previously collected \gls{snr} samples of a few specific locations within the area are known as prior. For the considered environment, 
We establish a novel framework for transmission control adaptation problem {for \gls{urllc}}, aiming to minimize the long-term average transmit power while guaranteeing the \gls{qos} requirement for all the devices within the target area. {Specifically, the \gls{qos} requirement is characterized as the \gls{dvp} from a cross-layer design.}
}

\item {{We employ a power scaling algorithm in conjunction with \gls{drl} algorithm for the transmission control adaptation with the help of the \gls{ckm}.}  Specifically, we first employ the \gls{gp} to obtain the \gls{ckm} of the overall target area and utilize an improved $K$-means clustering to divide the locations with similar channel statistics into clusters. Then we train a base DRL policy based on the location with known channel gain samples for each cluster. Finally, we conduct power scaling according to the \gls{ckm} within the same cluster via \gls{ckm}. Note that once all the base policies for the corresponding clusters are learned, the transmission control can achieve fast adaptation without re-training for a target area with accurate \gls{ckm}.}

\item {Then we utilize a meta-reinforcement learning algorithm that can quickly adapt to the new environment by learning from multiple location-specific environments incorporating the idea of \gls{maml} without the need for \gls{ckm}. The PPO-based meta-training algorithm is employed to train a policy that has good generalization capability among distinct location-specific environments. Then the meta-policy can adapt quickly to new location-specific environments with a few gradient update steps.}

\item {Simulation results show that the  PPO-based transmission control policy is efficient under various \gls{qos} requirements. 
Regarding the policy adaptation problem, it is also proved that the power scaling scheme and meta-\gls{drl} algorithm can guarantee the required \gls{qos} with 96.83\% and 99.55\% availability of the target area, respectively. The meta-\gls{drl} algorithm performs better than the power scaling scheme due to the interactions with new environments and adaptation. 
}
\end{itemize}

The remainder of this paper is organized as follows: Section~\ref{sec2} introduces the system model and presents the formulated problem. Section~\ref{sec3} presents the DRL-based transmission control policy for a stationary environment. Section~\ref{sec4} presents the power scaling algorithm in conjunction with the \gls{drl}-based algorithm via well-established \gls{ckm}. In Section~\ref{sec5}, a meta-reinforcement learning-based algorithm is presented. Comprehensive simulation results are presented in Section~\ref{sec6} and Section~\ref{sec7} concludes this paper.  Key notations in the system model and their definitions are summarized in Table~\ref{tab1}.

\section{System Model and Problem Formulations} \label{sec2}
We consider a downlink URLLC transmission scenario {for mission-critical \gls{iot} services} within a target area $\mathcal Z \in \mathbb R^{2}$  under the coverage of a \gls{bs}. 
The \gls{bs} engages in communication with {\gls{iot}  devices (e.g., industrial actuators, robots, smart grid controller) in the target area} and needs to determine the transmission parameters to meet stringent delay and reliability requirements for each device. For simplicity, the \gls{bs} and each device are assumed to have single antenna only, as our focus is on transmission control adaptation problem and learning; the presented methods can be generalized to multiple-antenna systems. It is assumed that the instantaneous \gls{csi} is unavailable at the \gls{bs} side before transmission due to, e.g.,  low latency constraints. Therefore, the \gls{bs} can only rely on statistical information of long-term channel gain samples collected from previously measured locations across the target area for transmission control. 

\begin{table}
\begin{center}
\caption{{ Notations and definitions}}
\begin{tabular}{|c | c|}
\hline
Notation & Definition \\
\hline
$\mathcal Z$  & Target area \\
$h(t)$  & Instantaneous channel coefficient at slot $t$\\
$p(t)$ & Instantaneous transmit power at slot $t$\\
$\gamma(t)$  &  Instantaneous channel gain  at slot $t$\\
$\Gamma(t)$  &  Instantaneous SNR       at slot $t$    \\
$\sigma_z^2$   & AWGN noise power                      \\
$R(t)$  &  Instantaneous coding rate  at slot $t$ \\
$A(t)$  &  Instantaneous arrival rate  at slot $t$ \\
$S(t)$  &  Instantaneous service rate  at slot $t$ \\
$Q(t)$  &  Queue length at slot $t$\\
$D(t)$  &  Delay experienced by the departure bits at slot $t$ \\
$\varepsilon$ & Outage probability for a single-shot transmission \\
$F_\gamma^{-1}(\varepsilon)$ & $\varepsilon$-quantile of $\gamma$ \\
$\mathcal D$  &  Channel gain data set\\
1 - $\xi$ &  Reliability target  \\
$D_{\rm max}$  &   Latency  target \\
\hline
\end{tabular}
\label{tab1}
\end{center}
\vspace{-0.4cm}
\end{table}

\subsection{Physical Layer Model}
The channel coefficient from the BS to an arbitrary device in the target area, denoted as $h$, is assumed to follow random block fading (comprising both small-scale and large-scale fading) drawn from an \emph{ unknown distribution}, conditioned on the location of the device denoted as  $\mbf{x}\in\mathcal{Z}$. 
The environment including propagation, blockages and scatterers is assumed to be static during the considered communication time period. Thus the channel distribution $f(h | \mbf{x})$ within $\mathcal Z$ exhibits spatial consistency and correlation. We consider time-slotted transmission with each slot consisting of $n\in \mathbb N$ channel uses. The channel coefficient remains constant within each time slot and varies independently from one slot to another.
The received signal at the {device} at slot $t$ is modeled  as 
\begin{equation}
    y(t) = \sqrt{p(t)} h(t) x(t) + z(t), 
\end{equation}
where $p(t)$ is the transmit power, $x(t)$ is the transmitted signal with unit power, and $z(t)$ is the additive white Gaussian noise (AWGN) with power $\sigma_z^2=BN_0$ i.e., $z\sim \mathcal {CN}(0, \sigma_z^2)$, where $B$ is the bandwidth of the channel and $N_0$ is the noise power spectral density.
The received \gls{snr} can thus be expressed as a random variable 
\begin{align}
 \Gamma(t)  = p(t) \gamma(t) \triangleq p(t) \frac{|h(t)|^2}{\sigma_z^2},
\end{align}
where $\gamma(t) \sim f( \gamma| \mathbf x) $ denotes the effective channel gain after normalization with the noise power at slot $t$ for location $\mathbf x$.

Due to the lack of instantaneous \gls{csi} at the \gls{bs}, {the effect of finite blocklength resulting from the noise instance is negligible. It is thus reasonable to work with the asymptotic outage probability with infinite blocklength~\cite{outage,yang2014}}. 
More specifically, we consider the $\varepsilon$-outage capacity, i.e., the maximum rate that the channel can support with success probability $1 - \varepsilon$ to model the transmission rate between the \gls{bs} and {device}, defined as:
\begin{equation}
     R_\varepsilon(t)= \sup\limits_{R} \left\{ R: \mathrm{Pr} \left(\log_2(1+\Gamma(t)) <R \right)<\varepsilon  \right\},
\end{equation}
and can be simplified as  
\begin{align}
    R_{\varepsilon}(t)=\log_2\left[1+p(t)\times  F^{-1}_{\gamma} (\varepsilon)\right],
    \label{outage rate}
\end{align}
where $ F^{-1}_{\gamma} (\varepsilon)$ is the $\varepsilon$-quantile of $\gamma$.

It is assumed that the explicit channel statistics of this target area are unknown to the transmitter. However, we assume that the \gls{bs} has a dataset $\mathcal D =\{\mbs { \gamma_d}, \mbf x_d\}_{d=1}^{M}$ collected over certain period from other devices across $M$ locations $\mbf x_d$ within the same target area $\mathcal Z$.  
Each $\mbs { \gamma_d}$ consists of $U$ previously collected channel gain samples at location $\mbf x_d$, i.e.,  $\boldsymbol{ \gamma_d}=[ \gamma_d^1,..., \gamma_d^U]$. {Here, we assume the channel process is stationary and ergodic, such that all channel gain samples at a given location are sourced from the same distribution and as long as the time duration for sample collection is long enough, these samples contain enough statistical features of the channel. } It is also assumed that the long-term statistics vary smoothly in space due to shared dominant paths, scatterers, etc. In addition, the location of each device is known at the \gls{bs} perfectly.  
\subsection{Link Layer Model}
\begin{figure}
    \centering
    \includegraphics[scale=0.53]{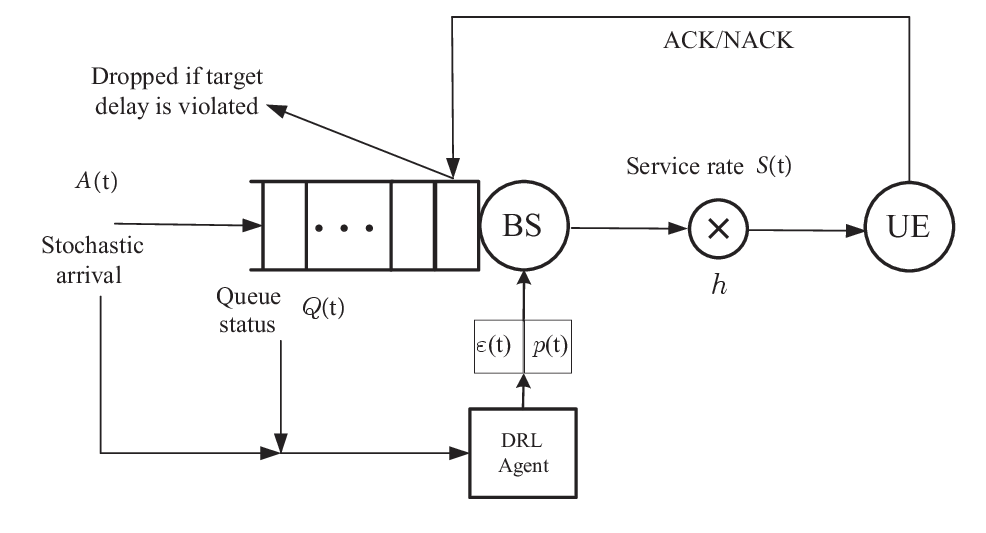}
    \caption{Queueing diagram of the cross-layer transmission model}
    \label{fig:linklayer}
    \vspace{-0.4cm}
\end{figure}

The link layer model depicted in Fig.~\ref{fig:linklayer} is similar to the one used in  \cite{ours}. The overall procedures are summarized as follows:
At the beginning of slot ${t \in \mathbb{N}}$, the data arrival rate and service rate are denoted as $A(t)$ and $S(t)$, respectively. {The traffic arrival process is modeled as a stochastic process with the average arrival rate $\lambda > 0$.}
The \gls{bs} first stores the arrived information bits in a first-in-first-out (FIFO) buffer. When each transmission is over, the {device} can reliably detect transmission errors and then send a one-bit error-free and delay-free acknowledgment (ACK) or negative acknowledgment (NACK) signal to inform the \gls{bs} whether the transmission was successful. { By neglecting these overheads and errors, we can gain an insight into the performance bounds. } Upon receiving the NACK signal, the transmission is said to be in outage and the \gls{bs} will retransmit the data. 
{The data in the buffer will be dropped when an ACK signal is received, or when the survival time of the data in the queue exceeds the target deadline, i.e., $D_{\rm max}$ time slots.}

According to the above mechanism, the service rate at slot $t$ is zero when an outage occurs. Thus the service rate 
(in bits per slot) can be expressed as
\begin{equation}
\begin{aligned}
 S(t)=
\begin {cases}
0, &\quad \quad \mathrm { Outage,} \\
n  R_{\varepsilon}(t),& \quad \quad \mathrm {No ~ outage,}
\end{cases}
\label{service-rate}
\end{aligned}
\end{equation}
where $R_{\varepsilon}(t)$ is the coding rate at slot $t$, $n$ denotes the channel uses in one slot, and 
\begin{align}
\mathrm{Pr}(\text{Outage}) %&= 1 - \mathrm{Pr}(\text{No outage}) \\
&= \mathrm {Pr}(\log_2(1+\Gamma(t)) < R_{\varepsilon}(t)).
\end{align}

At the end of slot $t$, and before outdated data is dropped, the temporary queue length is  given by
\begin{equation}\label{queueupdate}
Q_{\text{tmp}}(t+1)=\max \{Q(t)+A(t)-S(t), 0   \},
\end{equation}
where $Q(t)$ is the queue length at the beginning of slot $t$. 
{If the packet has survived $D_{\rm max}$ slots in the queue, the outdated data packet will be dropped.}
% Then we define the delay experienced by the data packet that arrived at  slot $t$ as $D(t)$,
% \begin{align}
% \mathbb D(t)  =\sum\limits_{i=1}^{t} \mathbb {I} \{D(i)> D_{\max} \} 
% \end{align}
% where $\mathbb I\{\cdot\}$ is the indicator function. 
To characterize the proactive outdated data dropping, we first  introduce  the accumulated number of arrival bits in the latest $D_{\max}$ slots at the end of slot $t$:
\begin{equation}
Q_{\text{th}}(t) = \sum\limits_{i=t-D_{\max}+1}^t A(i).
\end{equation}
Then we define the delay experienced by information bits that depart the queue at slot $t$ as $D(t)$, and the cumulative number of delay violation events can be characterized through queue length violation events {empirically} as 
\begin{align}
\mathbb D(t)  &=\sum\limits_{i=1}^{t} \mathbb {I} \{D(i)> D_{\max} \} \\ 
              &= \sum\limits_{i=1}^{t} \mathbb {I} \{Q_{\text{tmp}}(i+1)> Q_{\text{th}}(i) \},
\end{align}
where $\mathbb I\{\cdot\}$ is the indicator function. 

With the proactive outdated data dropping mechanism, the queue length  $Q(t+1)$ at the end of slot $t$ after the {outdated data dropping} step follows that 
\begin{equation} \label{droping}
Q(t+1)=\min \{Q_{\text{tmp}}(t+1), Q_{\text{th}}(t)\}.
\end{equation}

{We adopt the delay violation probability (DVP) as the performance metric related to reliability, which is defined as the probability that the actual delay $D(t)$ experienced by the bits depart the queue at slot $t$ exceeds the maximum tolerable $D_{\rm max}$ slots.  Clearly, according to the above queuing mechanism, the \gls{dvp} is equivalent to the queue length violation probability explicitly before outdated data dropping:}
\begin{align}
\mathrm {Pr} ({D(t)>D_{\max}})  &=\lim\limits_{T \to \infty} \frac{1}{T} \sum\limits_{t=1}^{T} \mathbb {I} \{D(t)> D_{\max} \} \\
                             &= \lim\limits_{T \to \infty} \frac{1}{T} \sum\limits_{t=1}^{T} \mathbb {I} \{Q_{\text{tmp}}(t+1)> Q_{\text{th}}(t) \},
\end{align}
herein, the \gls{dvp} is implicitly affected by the coding rate and the transmit power of each slot in the queueing system as in (\ref{queueupdate}) and (\ref{droping}).
\subsection{Problem Formulations}
{We aim to minimize the long-term average transmit power for the \gls{iot} devices while providing guaranteed reliability, i.e., bounded \gls{dvp}, with a stochastic arrival process across the entire target area.}
%where $\mathrm {DVP}(\mathbf {x})=\mathrm {Pr} ({D>D_{\max}})|\{h\sim f(h|\mbf x) \}$),  
% while minimizing the long-term average transmit power across the entire target area. 
%denoted as $\bar P(\mbf {x})=\lim\limits_{T\to \infty} {\frac{1}{T} \sum\limits_{t=1}^{T}  p(t) } | \{h\sim f(h|\mbf x)\}$ 
To tackle this problem, we first formulate the location-specific transmission control problem by optimizing the instantaneous transmit power and coding rate with known historical channel gain samples and then formulate a transmission control adaptation problem that aims to adapt to all the locations within the target area. 

For any given location $\mbf x$ with known historical channel gain samples, the power and rate control problem can be formulated as follows:
\begin{subequations}\label{P1}
\begin{align}
{\min\limits _{R(t),p(t)}} & \lim\limits_{T\to \infty}      {\frac{1}{T} \sum\limits_{t=1}^{T}  p(t) } \label{p1a}\\
{\text { s.t. }} \quad & {\mathrm {Pr}(D(t)>D_{\max})\leq \xi},   \label{p1b}
% \quad &  {0\leq p(t)\leq p_{\max}}, \label{p1c}
\end{align}
\end{subequations} 
where $p(t)$ and $R(t)$ are the transmit power and the coding rate at slot $t$, respectively. \eqref{p1b}  is the target  \gls{dvp} constraint. The objective  \eqref{p1a} is the long-term average transmit power.
% There are two concerns in the problem: The first is to determine the minimum transmit power with given queue length and instantaneous arrived information bits. The second one is between the coding rate and the outage probability, i.e., when transmit power is given, increasing the coding rate will increase the outage probability and more re-transmissions are likely needed and vice versa.  
Problem (\ref{P1}) is non-trivial thus we adopt a \gls{drl}-based approach to solve this problem which will be presented in detail in Section  \ref{sec3}. 

% Since this paper focuses on transmission control adaptation that can generalize to different channel distributions, 
Next, we generalize the location-specific problem (\ref{P1}) to account for all possible locations within the target area. We aim to find a transmission control adaptation policy $\Pi(\cdot|\mbf x)$  that can adapt to all the locations within the target area, including locations not in the dataset $\mathcal{D}$. We thus formulate the following problem: 
\begin{subequations}\label{P2}
\begin{align}
 \min\limits_{\Pi(\cdot|\mbf x)} \quad & \frac{1}{|\mathcal Z|} \int_{\rm x\in \mathcal Z} \bar P{(\bf x)}~d \mbf x \label{p2a}\\
{\text { s.t. }} \quad & {\mathrm {Pr}(D(t)>D_{\max}|\mbf x)\leq \xi, \quad \forall \mbf x \in \mathcal Z, }  \label{p2b}
%{\text { s.t. }} \quad & {\mathrm {DVP}(\bf x)\leq \xi, \quad \forall \bf x \in \mathcal R}.  \label{p2b}
\end{align}
\end{subequations} 
where $\bar P(\mbf {x})=\lim\limits_{T\to \infty} {\frac{1}{T} \sum\limits_{t=1}^{T}  p(t) } $ is the long-term average transmit power concerning the channel statistics at location $\mbf x$. Note that $\bar P(\mbf x)$ is not only related to the channel statistics of location $\mbf x$ but also related to the specified transmission control policy. Knowing the channel statistics can be utilized to determine the optimal transmission control policy and thus we omit the notation of the transmission control policy and simplify the notation as $\bar P(\mbf x)$. 
In problem (\ref{P2}), the optimization objective in (\ref{p2a}) is the minimum average transmit power over all the target area and the constraint in (\ref{p2b}) is the \gls{dvp} constraint for each location within the target area.  This problem is also non-trivial due to the following two aspects: 
 \begin{itemize}
\item   The \gls{dvp} constraints of locations without prior channel gain samples cannot be guaranteed perfectly due to the lack of statistical channel knowledge.
\item   The training time of the location-specific DRL algorithms for all the corresponding locations in $\mathcal D$ may be prohibitively long.
\end{itemize}

{We propose two solutions to solve the above policy adaptation problem (\ref{P2}). One is an offline power scaling approach in junction with \gls{drl} algorithm designed for (\ref{P1}) via a well-constructed \gls{ckm} without re-training. This solution works well for the environment with accurate \gls{ckm}. 
The other is a meta-reinforcement learning algorithm that requires limited interaction steps with the environment when \gls{ckm} is inaccurate or the new environment without \gls{ckm}.  The details shall be presented in Section \ref{sec4} and Section \ref{sec5}, respectively. }

% \section{\gls{drl}-Based Transmission Control Algorith} \label{sec3}
\section{Location-Specific Transmission Control } \label{sec3}
In this section, we employ a \gls{drl}-based algorithm to solve the power and rate control problem in \eqref{P1} {probabilistically} for any given location $\mbf x_d$ within dataset $\mathcal D$. 
% \textcolor{blue}
The policy requires selecting the transmit power $p(t)$ and  outage probability $\varepsilon(t)$ with the coding rate according to the estimated $\varepsilon$-outage capacity:
\begin{align}
    R_{\varepsilon}(t) = \log_2(1 + p(t)\widehat{F}_{\gamma}^{-1}(\varepsilon(t))),
\end{align}
{where $\widehat{F}_{\gamma}^{-1}(\varepsilon(t))$ is the estimated $\varepsilon$-quantile. Note that the traditional parametric approach for $\varepsilon$-quantile estimation usually assumes a parametric fading distribution, e.g., Rayleigh/Rician or Nakagami fading. This method is susceptible to model mismatch and thus influences the tail distribution of the outage probability for \gls{urllc}.  In this paper, we adopt a non-parametric method for the estimation based on historically collected channel samples in $\mathcal D$. }

\subsection{Non-parametric quantile estimation}
We employ a non-parametric estimator to estimate the channel gain quantiles of each location within the dataset as in \cite{srm}. For each location $\mbf x_d$ in the dataset $\mathcal D$, the $\varepsilon$-quantile of the channel gain, i.e., $\widehat{F}_{\gamma(\mbf{x}_d)}^{-1}(\varepsilon)$, can be estimated from the $U$ observations as:
\begin{align} \label{quantile}
    \hat{q}_{\varepsilon, d}=\mbs{\gamma}_{d,(r)},\ \ r=\lfloor U\varepsilon\rfloor,\ \ d=1,\ldots, M, 
\end{align}
where $\mbs{\gamma}_{d,(r)}$ is the $r$-th order statistics of $\mbs\gamma_d$ and $\lfloor\cdot\rfloor$ is the floor function that obtains the largest integer index smaller than $U\varepsilon$. To simplify the rate selection problem, we choose a discrete set of possible target outage probability levels, denoted $ \mathscr{E}=\{\varepsilon_1,..,\varepsilon_G\}$ with $G$ being the number of distinct outage probability levels. This generates a new data set $\mathcal D_{\mathscr{E}}=\{ \mbs{\hat q}_{ \mathscr{E}, d}, \mbs{\mathrm {x}}_d  \}_{d=1}^M$, where $\mbs {\hat q}_{\mathscr{E}, d} = [\hat q_{\varepsilon_1,d}, ..., \hat q_{\varepsilon_G,d}]$. It is worth mentioning that the quantile estimate does not require the distribution knowledge of the channl gain. Compared to parametric estimates, the main disadvantage is the excessive number of samples required for estimation when $\varepsilon$ is small. In fact, (\ref{quantile}) shows that $U$ scales inversely with $\varepsilon$, and $\hat{q}_{\varepsilon, d}$ is only well defined when $U\geq 1/\varepsilon$. However, due to the cross-layer transmission design allowing retransmissions, to achieve the target \gls{dvp} $\xi$, the $\varepsilon$-outage probability can be reduced by several orders of magnitude \cite{8490699}. Thus the number of channel gain samples can also be reduced.  % % \mbs {\vec \varepsilon}

\vspace{-0.2cm}
\subsection{MDP formulation and \gls{drl} algorithm}
For the convenience of analysis and operation, we first transform problem (\ref{P1}) as a \gls{mdp} and divide the \gls{mdp} into episodes each consisting of consecutive $T$ slots, where one slot corresponds to a step in the MDP. Specifically, we randomly choose one sample from the $U$ known channel gain samples in $\mbs {\gamma_d}$ as the channel coefficient in one slot to interact with the agent. {Since we aim to use the historical channel gain samples in an offline manner and thus we randomly choose the historical channel gain samples instead of randomly generating new channel gain samples.  } 
Then we define the framework of our RL problem:
\begin{itemize}
\item   {State space:} The state space $\mathcal S$ is composed of the queue length information and the number of the arrived information bits. Hence the state at slot $t$ is given by
\begin{align}
{s_t} = \{Q(t), A(t) \}.
\end{align}
\item  {Action space:} The action space $\mathcal A$ is composed of the outage probability and the transmit power. To ease the action tractability, we also divide the power into $H$ discrete levels, i.e., $\mathscr {P} = \{p_1,p_2,...,p_H\}$ with the help of the average channel gain through $\mathcal D$, thus the action at  slot $t$ can be expressed as
\begin{align}
{a_t} = \left\{ \varepsilon(t),  p(t) \right\}, \quad \forall \varepsilon(t) \in \mathscr{E}, \quad \forall p(t) \in  \mathscr P.
\end{align}
\item   {State transition function:} The state transition function $\mathcal P$ denotes a transition probability distribution from the current state $s_t$ to a new state $s_{t+1}$ after executing action $a_t$, i.e.,  $s_{t+1} \sim \mathrm {Pr}(~\cdot~|~s_t, a_t)$. 
\begin{align}
    \sum\limits_{s_{t+1}\in \mathcal S} \mathrm {Pr}(s_{t+1}|s_t, a_t) =1.
\end{align} 
{Note that the transition probability for a given state and an action is only related to the channel statistics in location $\mbf x_d$ and does not depend on the stochastic traffic arrival. This is because 
% To be more specific, the state transition randomness comprises of two aspects: The first is the stochastic traffic arrival in the new slot, which we do not need to consider since we assume we can observe this at the beginning of each slot. The second is the queue length update regarding the transmission outage. In this paper, we focus on the randomness regarding the transmission outage affected by the stochastic channel coefficient thus the state transition function is determined by the channel statistics of the current location. 
the stochastic traffic arrival in the new slot is observable at the beginning of each slot and thus can be considered deterministic. In this paper, we focus on the randomness regarding the transmission outage affected by the stochastic channel coefficient in each time slot. Thus the state transition function is determined by the channel statistics of the current location in a sense and there are two possible new states resulting from whether the transmission is in outage in (\ref{service-rate}). 
}

\item  {Reward function:}  The reward function $\mathcal R$ is critical for learning the optimal transmission policy. 
We propose a double-layer penalty reward to capture the queue length violation probability, or equivalently, \gls{dvp} as follows, 
\begin{equation}
r_t=
\begin {cases}
-p(t), &\quad \quad  Q_{\text{tmp}}(t+1) \leq Q_{\text{th}}(t)  \\
-p(t) - w(t) ,& \quad \quad Q_{\text{tmp}}(t+1) >Q_{\text{th}}(t)
\end{cases},
\end{equation}
\begin{align}
{
w(t)=
\begin {cases}
\Delta\left(\frac{\mathbb D(t)}{T\xi^*}\right)^{\nu}, &\quad \quad  \mathbb D(t) \leq T\xi^* \\
\Delta,& \quad \quad \mathbb D(t) >T\xi^*
\end{cases},
}
\end{align}
where $\Delta $ represents a significantly large penalty term, $T\xi^*$ denotes the target number of delay violation events which ensures the \gls{dvp} and $\nu$ is a positive integer that controls the increasing speed of $w(t)$ as a function of $\mathbb D(t)$. 
Specifically, the first layer penalty depends on the latency constraint,  once the equivalent queue length is violated, the penalty term emerges. However, only focusing on this penalty will result in the agent trying to not violate the target instead of ensuring $\mathbb D(t)$ less than the target number $T\xi^*$. Therefore, the second layer penalty term located in $w(t)$, which corresponds to the number of delay violations, is proposed.   
%When $\mathbb D(t)$ is small,  $w(t)$ is very small. As $\mathbb D(t)$ increases, the penalty term increases faster to address delay violation events. When $\mathbb D(t)$ approaches the violation target  $T\xi^*$, $w(t)$ holds a large value $\Delta$. As a result, the penalty term $w(t)$ encourages the agent to violate the delay threshold appropriately when the number of cumulative delay violations $\mathbb D(t)$ is small, while not violating the delay threshold when $\mathbb D(t)$ approaches the target number $T\xi^*$.
\end{itemize}

{Then we employ \gls{ppo} \cite{PPO} to solve the formulated problem due to its simplicity, sample efficiency, stability, and scalability. }
PPO is a policy gradient algorithm derived from the trust region policy optimization (TRPO) algorithm \cite{TRPO}. It can guarantee monotonically non-decreasing performance by optimizing the policy within the divergence constraint (named trust region). 

We first denote the objective function of a policy $\pi_{\theta}$ parameterized by $\theta$ as
\begin{align}
J(\theta)=\mathbb{E}_{s_0}\left[ V_{\pi_\theta}(s_0)\right]=\mathbb{E}_{\pi_\theta} \left[\sum_{t=0}^{T} \gamma_{\rm DIS}^tr(s_t,a_t)\right],
\end{align}
where $\gamma_{\rm DIS}$ is the discount factor.

{Suppose the old policy $\pi_{\theta_{\text{old}}}$ is parameterized as $\theta_{\text{old}}$.
The advantage function which captures how better the current action is than the average policy can be given as 
% we aim to find  a better policy $\pi_{\theta}$ such that $J(\theta) \geq J(\theta_{\text{old}})$. Then we can formulate the objective function of $\pi_{\theta_{\text{old}}}$ as the expectation form of  $\pi_{\theta}$ as follows,
}
% \begin{align}
%     J(\theta_{\text{old}})&=\mathbb{E}_{s_0}\left[ V_{\pi_{\theta_{\text{old}}}}(s_0)\right]\\
%     &=\mathbb{E}_{\pi_{\theta}} \left[\sum_{t=0}^{T} \gamma^t V_{\pi_{\theta_{\text{old}}}}(s_t) - \sum_{t=1}^{T} \gamma^t V_{\pi_{\theta_{\text{old}}}}(s_t) \right]\\
%     &=-\mathbb{E}_{\pi_{\theta}}\left[ \sum_{t=0}^{T}\gamma^t (\gamma V_{\pi_{\theta_{\text{old}}}}(s_{t+1}) - V_{\pi_{\theta_{\text{old}}}}(s_{t}) )\right].
% \end{align}
% Then we can derive the objective difference between new policy $\pi_{\theta}$ and old policy $\pi_{\theta_{\text{old}}}$
% \begin{align}
%     J(\theta) - J(\theta_{\text{old}}) &=\mathbb{E}_{s_0}\left[ V_{\pi_{\theta}}(s_0)\right]- \mathbb{E}_{s_0}\left[ V_{\pi_{\theta_{\text{old}}}}(s_0)\right]\\
%     &= \mathbb E_{\pi_{\theta}}\left[\sum_{t=0}^{T}\gamma^t  \Big [ r(s_t,a_t) +  \right.  \nonumber \\ 
%     & ~~~  \left.  \gamma V_{\pi_{\theta_{\text{old}}}}(s_{t+1}) - V_{\pi_{\theta_{\text{old}}}}(s_{t})\right]  \Bigg ]\\
%     &= \mathbb E_{\pi_{\theta}}\left[\sum_{t=0}^{T}\gamma^t  A^{\pi_{\theta_{\text{old}}}}(s_t,a_t) \right] \\
%     &= \sum_{t=0}^{T}\gamma^t \mathop \mathbb E\limits_{s,a \sim \pi_{\theta }} [A^{\pi_{\theta_{\text{old}}}}(s_t,a_t)]\\
%     &=\frac{1}{1-\gamma} \mathop \mathbb E\limits_{s,a \sim \pi_{\theta}} [A^{\pi_{\theta_{\text{old}}} }(s_t,a_t)]
% \end{align}
\begin{align}
A^{\pi_{ \theta_{\text{old}} } }(s_t,a_t) =r(s_t,a_t) +\gamma_{\rm DIS} V_{\pi_{\theta_{\text{old}}}}(s_{t+1}) - V_{\pi_{\theta_{\text{old}}}}(s_{t}). 
\end{align}
To guarantee the new policy $\pi_{\theta}$ performs better than the old policy $\pi_{\theta_{\text{old}}}$, we have to ensure $ \mathbb E_{s,a \sim \pi_{\theta}} [A^{\pi_{\theta_{\text{old}}} }(s_t,a_t)] \geq 0$. 
It is worth mentioning that the trajectory sampled from $\pi_{\theta_{\text{old}}}$ cannot be utilized to update $\pi_{\theta}$ directly. To improve sample efficiency, importance sampling is proposed to deal with the new policy by utilizing the advantage functions obtained from the old policy. The policy objective substitute is optimized as follows, 
\begin{align}
\mathop{\mathrm {max}}\limits_{\theta} ~ L(\theta) =  \mathop \mathbb E\limits_{s,a \sim \pi_{\theta_{\text{old}}}} \left[\frac{\pi_{\theta}(a_t|s_t)}{\pi_{\theta_\text{old}}(a_t|s_t)}A^{\pi_{\theta_{\text{old}}}}(s_t,a_t) \right],
\end{align}
with  Kullback-Leibler (KL) divergence constraint $\mathbb E [D_{\rm KL}[\pi_\theta||\pi_{\theta_{\text{old}}}]]\leq D_{\text {tar}}$ 
between the new and old policies.

Nevertheless, the complicated KL divergence constraint involved in TRPO makes it computationally inefficient and difficult to scale up for large-scale problems. To address this problem, PPO puts the KL divergence constraint into the objective function and adopts a clipping mechanism that allows it to use a first-order optimization and thus can reduce the computational complexity significantly. 
The optimization problem in PPO is shown in \eqref{PPOP} and the stochastic gradient ascent algorithm is utilized to update the network. 
\begin{figure*}
\begin{align}
\mathop{\mathrm {max}}\limits_\theta ~ L(\theta) =  \mathop \mathbb E\limits_{s_t,a_t \sim \pi_{\theta_{\text{old}} }} \left[ \min \left( \frac{\pi_{\theta}(a_t|s_t)}{\pi_{\theta_{\text{old}}}(a_t|s_t)},  \text{clip}\left(\frac{\pi_{\theta}(a_t|s_t)}{\pi_{\theta_{\text{old}}}(a_t|s_t)},1-\epsilon,1+\epsilon\right)  \right) A^{\pi_{\theta_{\text{old}}}}(s_t,a_t) \right], \ \text{clip}(x,a,b) = \begin{cases} x, & a < x < b \\ a, & x \leq a \\ b, & x \geq b \end{cases}
\label{PPOP}
\end{align}
\hrulefill
\end{figure*}
The overall algorithm is presented in Algorithm~\ref{Alg1}.
\begin{algorithm}[!h]
    \caption{PPO-based Algorithm}
    \label{Alg1}
    \renewcommand{\algorithmicrequire}{\textbf{Input:}}
    \renewcommand{\algorithmicensure}{\textbf{Output:}}
    \begin{algorithmic}
        \REQUIRE Randomly initialized policy network parameters $\theta$ %and value network $\phi$ 
        \FOR{episode = $1,...,Z$}
        \STATE $\pi_{\theta_{\text{old}}}$ = $ \pi_{\theta}$
        \FOR{$t=1,...,T$}
        \STATE Utilize $\pi_{\theta_{\text{old}}}$ to choose action $a_t$ and interact with the environment;
        \STATE Execute action $a_t$ and receive next state $s_{t+1}$, reward $r_t$ 
        \STATE Collect transition ($s_t$, $a_t$, $r_{t}$, $s_{t+1}$) to replay buffer $\mathcal B$
        \STATE $s_t\leftarrow s_{t+1}$
        % Collect the set of trajectories with length $T$ steps.
        \IF{ $t \text{ mod } T_g = 0$ }
        \STATE Load the transition tuples $(s, a,r,s)$ from replay buffer $\mathcal B$
        \STATE Compute the advantage function $A_t$ based on $\pi_{\theta_{\text{old}}}$ 
        \STATE Update the policy network by maximizing the PPO-clip objective function in (\ref{PPOP}) as: $\theta\leftarrow\theta+\alpha\nabla_{\theta} L(\theta)$ \\
        \STATE Reset the replay buffer $\mathcal B$
        \ENDIF
        \ENDFOR
        % \FOR{$epoch = 1,...,M$}
        % \STATE Sample a random minibatch of $b$ transition tuples $(s, a,r,s)$ from buffer $B$
        % \STATE Compute the advantage function $A_t$ based on $\pi_{\theta_{\text{old}}}$ %and current value function $V_t$
        % \STATE Update the policy network by maximizing the PPO-clip objective function in (\ref{PPOP}) as: $\theta\leftarrow \theta+\alpha\nabla_{\theta} L(\theta);$
        % % \STATE Update the value function by regression in (\ref{PPOPcritic}) as: $\phi \leftarrow \phi - \alpha_c \nabla_{\phi} L(\phi) $
        % \ENDFOR
        \ENDFOR
    \end{algorithmic}
\end{algorithm}
\section{ Policy Adaptation via Channel Knowledge Map}\label{sec4}
In this section, we first establish a \gls{ckm} based on the idea in \cite{srm,10105152}. 
% \textcolor{blue}
{Then we propose a power scaling algorithm based on the \gls{ckm} and the trained \gls{drl} algorithm to adapt to the new location with different channel statistics without re-training. To mitigate computational complexity and improve the performance of the power scaling scheme, we also divide the locations with similar channel statistics into the same clusters for (\ref{P2}).} 
\subsection{ Channel Knowledge Map}
Based on the dataset $\mathcal D_{\mathscr{E}}$, we will focus on the prediction of the estimated $\mathscr{E}$-quantiles of the new location $\mbf x^*$ that is not contained in $\mathcal D$ across the target area $\mathcal Z$ in the following. 
Firstly, we should standardize the $\varepsilon$-quantiles within $\mathcal D_{\mathscr{E}}$ as: 
\begin{align}
    \hat{q}'(\mathbf{x}_{d})=(\hat{q}_{\varepsilon, d}-\bar{q})/s,
\end{align}
where $\bar q =\frac{1}{M}\sum\limits_{d=1}^{M} \hat{q}_{\varepsilon, d}$ and $s=\sqrt{\frac{1}{D}\sum_{d=1}^{M} (\hat{q}_{\varepsilon, d} -\bar q)^2}$ are the sample mean and standard deviation of the estimated $\varepsilon$-quantile $\hat{q}_{\varepsilon, d}$ in (\ref{quantile}) across the $M$ locations.  
Based on \cite{dist-theory}, the quantile estimates derived from order statistics are asymptotically Gaussian and thus 
we assume the observation model $\hat{q}'_{\varepsilon, d} = {q}'_{\varepsilon, d} + \zeta$,  where $\zeta$ is an independent Gaussian random variable with zero mean and variance $\sigma_\zeta^2$.
Furthermore, it is assumed that $q'$ is a Gaussian process \cite{GP},
where $\mbf q'(\mbf X)=\left[q'(\mbf x_1),..., q'(\mbf x_M) \right]$ at any finite subset of locations $\mbf X =[\mbf x_1, ..., \mbf x_M]$ is jointly Gaussian such that 
\begin{align}
 \mathbf{q}^{\prime}(\mathbf{X})\sim\mathcal{N}(\boldsymbol{\mu}(\mathbf{X}),\boldsymbol{\Sigma}_{\mathbf{XX}}).
\end{align}

Here, the mean vector $\mbs \mu (\mbf X) \in \mathbb R^M$ is defined in terms of mean function $[\mbs \mu (\mbf X)]_i = m(\mbf  x_i; \mbs \eta_m)$ for $i = 1,...,M$ parameterized by $\mbs \eta_m$. The elements of the covariance matrix $\boldsymbol{\Sigma}_{\mathbf{XX}}\in \mathbb R^{M \times M}$ are given by $\left[ \mbs \Sigma (\mbf X)\right]_{i,j}=k(\mbf x_i, \mbf x_j, \mbs \eta_k)$, where $k(\mbf x_i, \mbf x_j, \mbs \eta_k)$ is a symmetric kernel function parameterized by $\mbs \eta_k$. In the context of radio channels, the absolute exponential kernel
\begin{align}
k(\mbf{x}_{i},\mathbf{x}_{j};\mbs{\eta}_{k})=\sigma_{k}^{2}\exp\left(-\frac{\Vert\mbf{x}_{i}-\mbf{x}_{j}\Vert_{2}}{d_{c}}\right),
\end{align}
with parameters $\mbs \eta_k=(\sigma_k^2, d_c)$ is referred to as the Gudmundson correlations model, and has been widely applied along with a log-distance mean function to model how the average channel gain varies across space.  

Note that $\sigma_k^2$ is the variance of the Gaussian process and $d_c$ is the correlation distance, which tends to be in the order of the size of blocking objects when modeling shadow fading. As shown in \cite{srm}, the Gudmundson correlation model is also well suited to model the $\varepsilon$-quantile of the channel gain which is therefore adopted to characterize the spatial correlation of $\mbf q '$. For the mean, we use $m(\mbf x_i)=0$, as commonly done in Gaussian processes \cite{GP}.

The \gls{ckm} is constructed for a regular grid of $L$ locations simultaneously, denoted $\mbf X^* = [\mbf x_1^*,...,\mbf x_L^*]$. In order to predict the Gaussian process at locations $\mbf X^*$, we express the joint distribution of the noisy observations $\mbf {\hat q}' (\mbf X) \in \mathbb R^{M}$ and quantiles in the grid $\mbf { q}' (\mbf X^*) \in \mathbb R^{L}$
\begin{align}
 \begin{bmatrix}\mathbf{q}^{\prime}(\mathbf{X}^{\ast})\\\hat{\mathbf{q}}^{\prime}(\mathbf{X})\end{bmatrix}\sim\mathcal{N}\left(\mathbf{0},\begin{bmatrix}\boldsymbol{\Sigma}_{\mathbf{X}^{\ast}\mathbf{X}^{\ast}} &\boldsymbol{\Sigma}_{\mathbf{X}^{\ast}\mathbf{X}}\\\boldsymbol{\Sigma}_{\mathbf{XX}^{\ast}} &\boldsymbol{\Sigma}_{\mathbf{XX}}+\sigma_{\zeta}^{2}\mathbf{I}_{M}\end{bmatrix}\right), 
\end{align}
where $\mathbf{I}_{M} \in \mathbb R^{M\times M}$  is the identity matrix. The prediction distribution $\mathbf{q}^{\prime} (\mathbf{X}^{\ast})\vert\boldsymbol{\vartheta}$ is also a multivariate Gaussian distribution 
\begin{align}
E[\mathbf{q}^{\prime}(\mathbf{X}^{\ast})\vert\boldsymbol{\vartheta}]&=\boldsymbol{\Sigma}_{\mathbf{X}^{\ast}\mathbf{X}}(\boldsymbol{\Sigma}_{\mathbf{XX}}+\sigma_{\zeta}^{2}\mathbf{I}_{M})^{-1}\hat{\mathbf{q}}^{\prime}(\mathbf{X}), \label{GPmean}\\
\text{Cov}[\mathbf{q}^{\prime}(\mathbf{X}^{\ast})\vert\boldsymbol{\vartheta}]&=\boldsymbol{\Sigma}_{\mathbf{X}^{\ast}\mathbf{X}^{\ast}}-\boldsymbol{\Sigma}_{\mathbf{X}^{\ast}\mathbf{X}}(\boldsymbol{\Sigma}_{\mathbf{XX}}+\sigma_{\zeta}^{2}\mathbf{I}_{M})^{-1}\boldsymbol{\Sigma}_{\mathbf{XX}^{\ast}}, \label{GPvariance}
\end{align}
where $\vartheta=(\widehat{\mathbf{q}}^{\prime}(\mathbf{X}^{\ast}), \mathbf{X}, \mathbf{X}^{\ast}, \mbs \eta_k)$.
Equations (\ref{GPmean}) and (\ref{GPvariance}) constitute the predictive distribution for $q'(\bf X^*)$ and are referred to as the predictive mean and covariance, respectively. This model produces a full distribution of the predicted quantiles. For simplicity, we will directly utilize the mean value as the predicted $\varepsilon$-quantile regardless of the uncertainty of the estimates. 
{The main reason is two-fold: 
Firstly, when conducting the coding rate and transmit power in the virtual environment with estimated $\varepsilon$-quantiles and the true environment with uncertainty in the $\varepsilon$-quantiles, the difference between the two corresponding outage probabilities in a single time slot is quite small due to the increased outage probability caused by cross-layer design \cite{8490699}. 
Secondly, assuming the estimation error exists. When deploying the trained agent in the true environment with worse channel condition, more outage transmissions are likely to occur. As long as the queue length is long enough, the trained agent will adapt the transmission strategy to be more conservative, leading to a choice of increased coding rate and increased transmit power with a smaller outage probability, to ensure the delay violation probability; and vice versa. As a result, the above two mechanisms can alleviate the impact of the uncertainty in channel statistics.}
%This will result in minor overestimation or underestimation of the $\varepsilon$-quantile and thus will influence the corresponding outage probability of the transmission compared with the true value. However, the impact on the outage probability brought by the prediction error remains constant after the \gls{ckm} establishment, and thus this impact can be alleviated through the adaptive \gls{drl} agent according to the instantaneous queue length, which is presented in the Section \ref{sec3}.

\vspace{-0.3cm}
\subsection{Power Scaling Scheme based on \gls{ckm}}
In the previous section, an efficient PPO-based DRL algorithm has been utilized to solve the outage probability and power control problem in a specific location with historical channel gain samples. 
{Here in this subsection, we aim to provide \gls{qos} guarantee for all the locations in $\mathcal Z$ that do not necessarily have historical channel gain samples using the established \gls{ckm}.} %, consisting of those locations within $\mathcal D$ while without the corresponding DRL algorithms training and locations not in $\mathcal D$. 

Intuitively, the trained \gls{ppo} agent can find the optimal transmit power and outage probability in each \gls{mdp} step for the specific location while guaranteeing the \gls{qos} requirement. From the state transition function perspective, this trained agent cannot efficiently guarantee the \gls{qos} for a new location with different channel statistics.
%This optimality is introduced from the tradeoff between the $\varepsilon$-quantiles of $\gamma$ and the outage probability, i.e., when choosing a larger outage probability $\varepsilon$, the corresponding $\varepsilon$-quantile of $ \gamma$ is also larger.  
When directly deploying the trained agent to a new location with better channel conditions, the agent will underestimate the channel statistics and thus will consume more transmit power than needed while the QoS requirement is met. Conversely, when deploying the trained agent to a new location with worse channel conditions, the agent will overestimate the channel statistics and the QoS requirement is more likely violated. 

% \textcolor{blue}{Note that the only difference between two different locations within the target area is the channel statistics. This difference will result in different state transition probabilities and cannot guarantee the \gls{dvp} constraint if we adopt the same action for these two locations \gls{mdp} with the same initialized state.}
Thus we propose the following power scaling scheme to reduce the transmit power for the environment with better channel conditions and increase transmit power for the environment with worse channel conditions based on the established \gls{ckm}. For notation simplicity, we assume $\pi_s(s_t)=\{\varepsilon(t), p_s(t)\}$, where $\pi_s$ is the policy trained from the source domain, i.e., the location with known estimated quantiles $ \mbs {\hat q}_{\mathscr {E}, s}$, and $s_t$ is the current state observation regardless of the domain. Thus the coding rate at the source domain with state observation  $s_t$ can be expressed as 
\begin{align}
    R_s(t)=\log_2\left[1+p_s(t)\times \widehat F_{ \gamma_s}^{-1}(\varepsilon(t))\right], 
\end{align}
where $\gamma_s$ is the channel gain in the source domain. 
When deploying the policy $\pi_s$ in the target domain, i.e., new locations with different channel statistics, with state observation $s_t$, we directly input the state observation $s_t$ to the policy $\pi_s$. Then we just scale the output transmit power $p_s(t)$ according to the $\varepsilon$-quantile difference between the source domain and the target domain while keeping the outage probability unchanged as follows, 
\begin{align}
    p_t(t) =  p_s(t) \times \frac{ \widehat F_{ \gamma_s}^{-1}(\varepsilon(t))}{ \widehat F_{ \gamma_t}^{-1}(\varepsilon(t))},  \label{scaling}
\end{align}
where $\gamma_t$ denotes the channel gain in the target domain. 
This power scaling scheme can eliminate the channel statistics difference and thus guarantee that the state transition probability in the target domain is similar to the one in the source domain, i.e., similar \gls{dvp} performance in a sense. 
% Regarding the power scaling scheme in (\ref{scaling}), we can conclude the following proposition: 
\newtheorem{theorem}{Theorem}
\newtheorem{proposition}{Proposition}
\begin{proposition}
If the $\varepsilon$-quantiles of the source domain and the target domain satisfy $ F_{ \gamma_s}^{-1}(\varepsilon)=\kappa  F_{ \gamma_t}^{-1}(\varepsilon), ~ \forall \varepsilon \in \mathscr E $, $\kappa$ is a constant, the optimal policy trained from the source domain after power scaling is also optimal in the target domain. 
\end{proposition}

\begin{algorithm}[!t]
    \caption{The improved $K$-means clustering Algorithm}
    \label{Alg2}
    \renewcommand{\algorithmicrequire}{\textbf{Input:}}
    \renewcommand{\algorithmicensure}{\textbf{Output:}}
    \begin{algorithmic}
        \REQUIRE  Predictive quantiles in the grid $q_{\mathscr E}'(\bf X^*)$,   the  number of clusters $K$  
        \STATE Randomly choose $K$  samples $q_{\mathscr E}'(\mathbf x^*_k), 1\leq k\leq K$ from  $q_{\mathscr E}'(\bf X^*)$ as central direction $\mu_k=q_{\mathscr E}'(\mathbf x^*_k)$
        \REPEAT
        \FOR{$j$ = 1,...,$L$}
        \STATE Compute cosine similarity between sample $q_{\mathscr E}'(\mathbf x^*_j)$ and each central direction $\mu_k=q_{\mathscr E}'(\mathbf x^*_k)$:\\ \quad \quad  \quad \quad \quad \quad  $s_{jk}=\frac{q_{\mathscr E}'(\mathbf x^*_j) \cdot q_{\mathscr E}'(\mathbf x^*_k)}{\Vert q_{\mathscr E}'(\mathbf x^*_j) \Vert_2  \times \Vert q_{\mathscr E}'(\mathbf x^*_k) \Vert_2 }$
        \STATE Divide $ q_{\mathscr E}'(\mathbf x^*_j) $ into cluster $C_\chi$ as:\\  \quad \quad  \quad \quad \quad \quad  $\chi= \mathop{\arg \max}\limits_{k=1,...,K} s_{jk}$
        \ENDFOR
        \FOR{$k$ = 1,...,$K$}
        \STATE Compute the new central direction of each cluster as:\\ \quad \quad \quad \quad $\mu'_k = \frac{1}{|C_k|} \sum_{q_{\mathscr E}'(\mathbf x^*_i)\in C_k} \frac{q_{\mathscr E}'(\mathbf x^*_i)}{\Vert q_{\mathscr E}'(\mathbf x^*_i) \Vert_2}$
        \STATE \textbf{if} $\mu'_k \neq \mu_k$  \textbf{do}
        \STATE \quad $\mu_k=\mu'_k$
        \STATE \textbf{else}
        \STATE \quad $\mu_k=\mu_k$
        \STATE \textbf{end if}
        \ENDFOR
        \UNTIL The central direction of each cluster remains constant
        \ENSURE  $\{C_{1}, C_{2},..., C_{K}\}$
    \end{algorithmic}
\end{algorithm}

\begin{proof}
% In the source domain, the action is $a_t=\pi_s(s(t))=\{\varepsilon(t), p(t)\}$,
The coding rates in the source domain and the target domain are 
\begin{align}
    R_s=&\log_2[1+p_s(t)  F_{ \gamma_s}^{-1}(\varepsilon(t))],\label{Rs} \\ 
    R_t=&\log_2[1+p_t(t)  F_{ \gamma_t}^{-1}(\varepsilon(t))], \label{Rt}
\end{align}
respectively. By substituting $ F_{\gamma_t}^{-1}(\varepsilon)=\frac{1}{\kappa} F_{ \gamma_s}^{-1}(\varepsilon)$ in to (\ref{Rt}), we can obtain the coding rate in the target domain as 
\begin{align}
    R_t=&\log_2[1+\frac{p_t(t)}{\kappa}  F_{ \gamma_s}^{-1}(\varepsilon(t))]\\
       =&\log_2[1+\rho(t) F_{ \gamma_s}^{-1}(\varepsilon(t))], \label{subs}
\end{align}
where $\rho(t)=\frac{p_t(t)}{\kappa}$.
Based on (\ref{subs}), we can find that the state transition function of the target domain and the source domain are quite similar and the only difference lies in the parameter $\kappa$, which remains constant for all the possible $\varepsilon$. 
% To minimize $\frac{1} \sum_{t=1}^{\infty} \rho(t)$ in the target domain, the optimal policy in the source domain can be directly utilized.  
% Thus, the objective in the target domain $\frac{1}{T} \sum_{t=1}^{\infty} p_t(t)$ is equivalent to $ \frac{1}{\kappa} \frac{1}{T} \sum_{t=1}^{\infty} p_t(t) =  \frac{1}{T} \frac{1}{\kappa}\sum_{t=1}^{\infty} p_t(t) = \frac{1}{T} \sum_{t=1}^{\infty} \rho(t) $
% For the source domain, the minimal average transmit power obtained from $\pi_s^*(\cdot)$ is denoted as $\frac{1}{T}\sum_{t=1}^{\infty} p_s(t) \Leftrightarrow \frac{1}{T} \sum_{t=1}^{\infty} \rho(t) \Leftrightarrow \frac{1}{T} \frac{1}{\kappa}\sum_{t=1}^{\infty} p_t(t)$. 
% \textcolor{blue}
 It can be easily proved that minimizing the average transmit power $p_t(t)$ with channel distribution $ \gamma_t \sim F_{ \gamma_t}(x)$ in (\ref{Rt}) is equivalent to minimizing the average transmit power $\rho(t)$ with channel distribution $ \gamma_s \sim F_{ \gamma_s}(x)$ in (\ref{subs}) by decoupling the constant parameter $\kappa$. Thus the power scaling scheme shown in (\ref{scaling}) is optimal when the condition $F_{ \gamma_s}^{-1}(\varepsilon)=\kappa F_{ \gamma_t}^{-1}(\varepsilon), ~ \forall \varepsilon \in \mathscr E$ holds.
% On the other hand, it can be easily proved that if the scaled optimal source domain policy $\pi_s^*(\cdot)$ is not optimal in the target domain, when utilizing the optimal policy $\pi_t^*(\cdot)$ into the source domain will have better performance, which contradicts with the fact. 
\end{proof}

Practically, most of the channel statistics across space can not meet the requirement $F_{ \gamma_s}^{-1}(\varepsilon)=\kappa F_{ \gamma_t}^{-1}(\varepsilon), ~ \forall \varepsilon \in \mathscr E$ strictly. There will be a mismatch among different channel distributions. To reduce computational complexity and improve the performance of the power scaling scheme, we adopt an improved $K$-means clustering to divide the $L$ $\mathscr E$-quantiles of the considered grid area into $K$ clusters. The improved $K$-means clustering can divide the most similar $\mathscr E$-quantiles by cosine similarity into the same cluster, i.e., ensuring $\widehat F_{ \gamma_s}^{-1}(\varepsilon)\approx\kappa \widehat F_{ \gamma_t}^{-1}(\varepsilon)$,  within the target area and thus reduce the power consumption resulting from channel distribution mismatch. The details of the improved $K$-means clustering algorithm are shown in Algorithm~\ref{Alg2}. {Specifically, we first randomly choose $K$ $\mathscr E$-quantiles as the central direction of the $K$ clusters. Then we can compute the cosine similarity of all the other $L-K$ $\mathscr E$-quantiles with respect to the $K$ central direction and divide the corresponding $\mathscr E$-quantiles into different clusters.  Finally, we update the central direction of each cluster by averaging the normalized direction of all the $\mathscr E$-quantiles within each cluster. This algorithm terminates until the central direction of each cluster remains constant. }
After determining the $K$ clusters, we need to choose $K$ base locations from those clusters as source domains and train $K$ agents from the $K$ base locations respectively to ensure the effectiveness of the power scaling scheme. Once the $K$ base policies are trained, the transmission control for any given location within the target area $\mathcal Z$ can be achieved without the need for any additional training time.
%Finally, we train $K$ different agents from $K$ base locations respectively. 

\section{Transmission Control Adaptation via Meta-DRL}\label{sec5}
In the previous section, we have proposed a power scaling scheme that can provide \gls{qos} guarantee for the user located in the target area $\mathcal Z$ with the help of the \gls{ckm}. However, the \gls{ckm} may be inaccurate since it is just a predicted map and we do not know the realistic channel statistics indeed. Thus the power scaling scheme may not work well when the predictive quantiles of the target domain are inaccurate. 
In this section, we propose a meta-reinforcement learning algorithm for (\ref{P2}) based on \gls{maml}\cite{maml} framework from a different perspective. 

The core idea of MAML is to pre-train meta parameters on a large variety of different tasks and quickly adapt to a new task based on the meta parameters after a few step gradient updates. 
The reinforcement learning framework based on \gls{maml}, as outlined in \cite{maml}, is directly applicable to our problem, given its emphasis on addressing on-policy MDP.  Notably, the PPO algorithm is recognized as an on-policy reinforcement algorithm, which aligns with the scope of this approach. In the following, we will first introduce the elements of meta-reinforcement learning and then introduce the \gls{maml}-based reinforcement learning algorithm.

\vspace{-0.2cm}
\subsection{Meta-Reinforcement Learning Elements}
The objective of meta-reinforcement learning is to find a model that can quickly adapt to new tasks comprising similar tasks in different environments or different tasks in the same environment. In our problem, the former is considered. We first define a meta-task set $\mathcal N$ consisting of $|\mathcal N|=N$ tasks. Each task $\mathcal T_i,~  i \in \mathcal N $, denotes a long-term power minimization problem in (\ref{P1}) for a specific location with known channel gain samples. Thus $\mathcal T_i$ can be modeled as an MDP, which is similar to the MDP expression in Section~\ref{sec3}. The key elements of the MDP also include the state space $ \mathcal S$, action space $\mathcal A$, state transition function $\mathcal P$ and reward function $\mathcal R$. 
{Note that the action space in this MDP is reformulated as the coding rate and the transmit power directly without the need for channel gain samples and the corresponding $\varepsilon$-quantiles in advance. 
More specifically, the known channel gain samples of this given location will be utilized to interact with the agent for task $\mathcal T_i$.
The main objective of this agent is to appropriately determine the current coding rate and the transmit power by interacting with the environment while guaranteeing the strict \gls{qos} requirement.}
The tasks in the meta-task set are selected according to the channel quality of the $M$ locations with known channel gain samples to ensure the meta-learner can learn the policy with good generalization capability under various environments.

\vspace{-0.3cm}
\subsection{ Meta-Reinforcement Learning Algorithm}
Meta training is a critical procedure for meta-reinforcement learning to obtain a good network initialization parameter. 
In the following, we will present the meta-reinforcement learning algorithm based on \gls{maml} algorithm. The training phase is divided into two levels, i.e., inner-level and outer-level. The inner-level individually optimizes the network parameter on each task gradually. The outer-level aims to update the global parameters with the updated inner-level parameters on the meta-task set. 
At the beginning of the meta-training phase, each task inherits the global initialized network parameters and then performs inner-level update to update its parameter. The updated parameter of each task will contribute to the outer-level update. 
It is worth mentioning that the inner-level update can be viewed as a normal update of the previous PPO-based algorithm which aims to find the optimal coding rate and the transmit power for the specified task with a notable difference that here we assume the coding rate and the transmit power are continuous variables instead of discrete variables. 
In the previous section, we have presented the details of the network update and these formulas can be directly utilized in this part\footnote{It is worth mentioning that although the action space types of these two problems are different, \gls{ppo} is an effective algorithm that can handle both discrete action space and continuous space. The core idea of \gls{ppo} algorithm is the same while the only difference is that the discrete-\gls{ppo} utilizes softmax as output action distribution and continuous-\gls{ppo} adopts normal distribution or beta distribution as the distribution of the action. In this paper, we adopt beta distribution for meta-learning action due to the property of the action space elements being all positive and bounded.}.  The details of the inner-level update and outer-level update are presented in the following. 
\begin{algorithm}[!t]
    \caption{\gls{maml}-based \gls{drl} Algorithm}
    \label{Alg3}
    \renewcommand{\algorithmicrequire}{\textbf{Input:}}
    \renewcommand{\algorithmicensure}{\textbf{Output:}}
    \begin{algorithmic}
        \REQUIRE Randomly initialized policy network parameters $\theta$ % and value network $\phi$ 
        \FOR{episode = $1,...,Z$}
        % \FOR{$t=1,...,T$}
        \FOR{ $\forall \mathcal T_i \in \mathcal T$}
        \STATE Set $\pi_{\theta_i}=\pi_{\theta}$
        \STATE Update the inner-level policy network $\theta_i$ based on (\ref{PPOP}) as $\theta_i \leftarrow \theta_i+\alpha\nabla_{\theta_i} L(\theta_i)$ with every $T_g$ time steps by Algorithm ~\ref{Alg1}
        \STATE Collect $T$ continuous transition tuples using the updated parameter $\theta_i$ to replay buffer $\mathcal B'_i$
        % \STATE Collect the transition consists of $B$ tuples with the updated policy network parameter 
        % \STATE Collect $M$ trajectories $\mathcal Y=\{\tau_i^1,...,\tau_i^M\}$ using $\pi_\theta$
        % \STATE Update the inner-level policy network $\theta_i$ based on (\ref{PPOP}) as $\theta_i \leftarrow \theta_i+\alpha_a\nabla_{\theta_i} L(\theta_i);$
        % \STATE Update the inner-level critic network $\phi^i$ based on (\ref{PPOPcritic}) as  $\phi_i \leftarrow \phi_i - \alpha_c \nabla_{\phi_i} L(\phi_i) $
        % \STATE Collect $M$ trajectories $\mathcal U'_i=\{{\tau'}_i^1,...,{\tau'}_i^M\}$ using $\pi_{\theta_i'}$
        \ENDFOR
        \STATE Update the outer-level policy network parameter $\theta$ as in (\ref{meta-para}) %and critic network parameter $\phi$ as in (\ref{meta-para})
        % \ENDFOR
        \ENDFOR
    \end{algorithmic}
\end{algorithm}

For each task $\mathcal T_i$, the agent interacts with the environment based on the initialized meta policy $\pi_{\theta_i}=\pi_{\theta}$. The policy network parameter $\theta_i$ is updated as shown in Algorithm~\ref{Alg1} every $T_g$ time steps. 
When the current episode terminates, we employ the updated network to interact with the environment for $T$ time steps and store the trajectory into the replay buffer $\mathcal B'_i$. 
Once the trajectories of the updated policy networks for all the tasks in $\mathcal N$ are collected, we can compute the policy gradient to update the outer-level policy network parameter $\pi_\theta$. The outer-level policy network $\theta$  can be updated as follows: %and critic network parameter $\phi$ 
\begin{align}
% {
% \begin{cases}
    \theta\leftarrow \theta + \beta \nabla_\theta \sum\limits_{i=1}^N  L(\theta_i, \mathcal B_i') 
    % \phi  \leftarrow  \phi - \beta_c \nabla_\phi \sum\limits_{i=1}^N  L(\phi_i, \mathcal B_i')
% \end{cases}
% },
\label{meta-para}
\end{align}
where $\beta$ is the learning rate of the meta-policy, $L(\theta_i, \mathcal B_i') $ is the loss function with respect to $\theta_i$ and calculated from the new trajectory in $\mathcal B_i'$ as in (\ref{PPOP}). In the next episode, the updated meta parameter $\theta$ will be employed as a new initialization for each task $\mathcal T_i$. The overall procedures are summarized in Algorithm~\ref{Alg3}.

By iteratively executing the inner-level update and the outer-level update, a meta-policy for transmit power and coding rate control can be trained among environments with different channel statistics.  Once the meta-policy is trained, the meta-policy can be utilized as an initialization and then optimized for the new task by only executing a few gradient updates.
\subsection{Deployment and Overhead  of the Meta-DRL algorithm} 
% The PPO and meta-DRL models are trained offline, which means that the models are trained during the idle time of the central controller using the trajectories and the corresponding rewards sampled in historical tasks. Hence, we ignored the time and energy consumption of the training process of the proposed algorithms. In actual VR applications, we first train the  model offline. The trained model is considered as an initialization model for new learning tasks and can be directly implemented without training. Meanwhile,
% when implementing a new task, the proposed meta learning algorithm can further improve its pre-trained initialization model to achieve better reliability of the studied network using a small number of iterations.

{The PPO and meta-DRL models are trained offline. More specifically,  they are trained based on the virtual environment consisting of the pre-collected historical channel gain samples in dataset $\mathcal D$. The training utilizes the trajectories and the corresponding rewards sampled from the interactions between the agent and the virtual environment before the \gls{drl} algorithm deployment. Hence, we can ignore the time consumption of the training process. 
When deploying the DRL algorithm in a practical setup, we first train the proposed two models offline. The trained PPO-based model can be directly implemented without training but with the requirement of the clustered CKM. The trained meta-DRL model can be regarded as an initialization model for new learning tasks and can be directly implemented. After a small number of gradient update steps, the meta-DRL model can be further improved in a new environment rather quickly. 
% When deploying the \gls{drl} algorithm to the practical scenario, we first train the proposed two models offline, respectively. The trained PPO-based model can be directly implemented without training with the requirement of the clustered \gls{ckm}. The trained meta-\gls{drl} model can be regarded as an initialization model for new learning tasks and can be directly implemented. After a small number of gradient update steps, the meta-\gls{drl} model can be further improved quickly in a new environment.
% Meanwhile, when implementing a new task, the proposed meta-learning algorithm can further improve its pre-trained initialization model to achieve better reliability using a small number of gradient update steps. 
}

% \newpage
\section{Simulation Results}\label{sec6}
% \begin{figure}[h]
% \vspace{-0.2cm}
% \centering
% \subfigbottomskip=-2pt %两行子图之间的行间距
% \subfigcapskip=-5pt %设置子图与子标题之间的距离
% \subfigure[Learning curves of the PPO algorithm with 3 random seeds of the environment, target \gls{dvp} $\xi=10^{-3}$ and $D_{\rm max}  =5 $ slots.]
% {\includegraphics[width=0.8\linewidth]{Figure/Traning-curves-r.eps}\label{fig:training}}
% \subfigure[Impact of reliability and latency.]
% {\includegraphics[width=0.8\linewidth]{Figure/tradeoff.eps}\label{fig:tradeoff}}
% \caption{Convergence curves of the PPO-based algorithm and latency, reliability tradeoff.}
% \vspace{-0.5cm}
% \end{figure}

In this section, we first validate the effectiveness of the  \gls{ppo}-based DRL algorithm for a stationary environment under various \gls{qos} constraints. Then we establish a synthetic scenario that constitutes propagation conditions according to standard 3GPP NR Urban Micro-Cell scenario with NLoS models generated by the simulation tool QuaDRiGa \cite{quadriga} for numerical evaluation. 

\vspace{-0.2cm}
\subsection{DRL-based Transmission Control Algorithm Validation} \label{Simu-a}
In this subsection, we will validate the effectiveness of the  PPO-based algorithm for the problem (\ref{P1}). For simplicity, we utilize the Rayleigh fading channel with unit variance. {We use Poisson process to model the stochastic traffic arrival process and the average arrival rate is $\lambda=800$ bits/slot.} The action space consists of $H=18$ power levels and $G=20$ outage probability levels. The peak power is set as $p_{\rm max}=-\frac{2^{\lambda/n}-1}{\log[1 - (\xi^{D_{\rm max}^{-1}}/D_{\rm max})]}$ empirically. The concrete power levels are assembled as $\mathscr P=\{\frac{1}{H}p_{\rm max}, \frac{2}{H}p_{\rm max},..., p_{\rm max} \}$ and the outage probability levels are assembled as $\mathscr{E}=\{\frac{1}{100}, \frac{2}{100},..., \frac{20}{100}\}$. 
To capture the \gls{dvp} more accurately, we assume each episode comprises $10/\xi$ steps and there are $100$ episodes for training. Regarding PPO, we adopt PPO-clip with a clip ratio $\epsilon=0.2$. The reward discount factor $\gamma_{\rm DIS}=0.99$. 
%generalized advantage estimation (GAE) parameter $\lambda_{\rm GAE} = 0.95$ and the 
Both the actor-network and critic-network learning rates are set to $l_r^a=l_r^c=\alpha=2\times 10^{-4}$.  $T_g$ is set as $2000$ and the minibatch size is $b=128$.  {The actor and critic networks are fully connected deep neural networks which have two hidden layers with 128 neurons, and the Tanh function is used as activation. }

% \begin{figure}[t]
% % \vspace{-0.5cm}
%     \centering
%     \includegraphics[scale=0.58]{Figure/Traning-curves-r.eps}
%     \caption{\textcolor{blue}{Learning curves of the PPO algorithm with 3 random seeds of the environment, target \gls{dvp} $\xi=10^{-3}$ and $D_{\rm max}  =5 $ slots}}
%     \label{fig:training}
%     \vspace{-0.5cm}
% \end{figure}

\begin{figure}[t]
% \vspace{-0.5cm}
    \centering
    \includegraphics[scale=0.58]{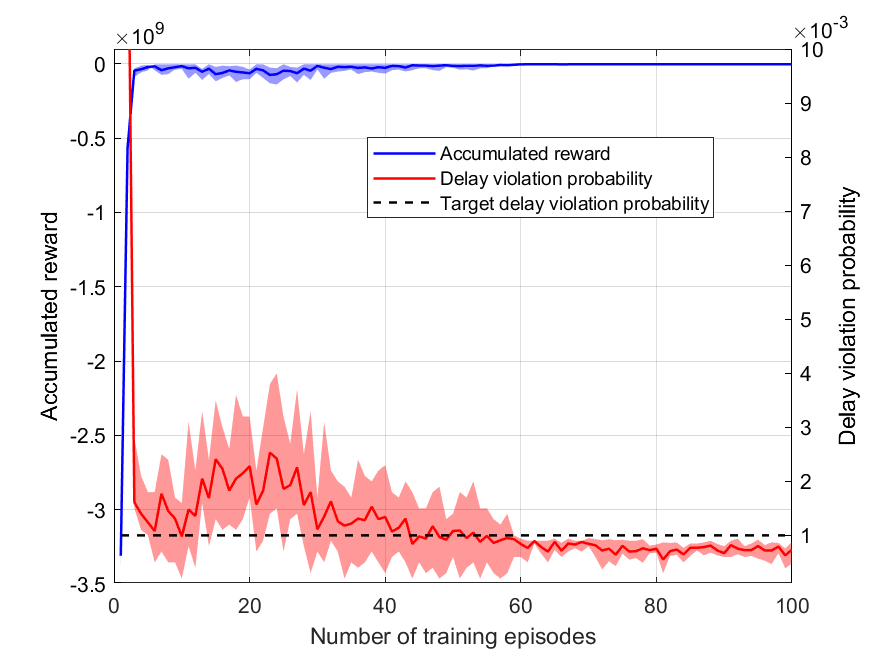}
    \caption{\textcolor{blue}{Learning curves of the PPO algorithm, target \gls{dvp} $\xi=10^{-3}$ and $D_{\rm max}  =5 $ slots}}
    \label{fig:training}
    \vspace{-0.5cm}
\end{figure}

\begin{figure}[t]
% \vspace{-0.5cm}
    \centering
    \includegraphics[scale=0.58]{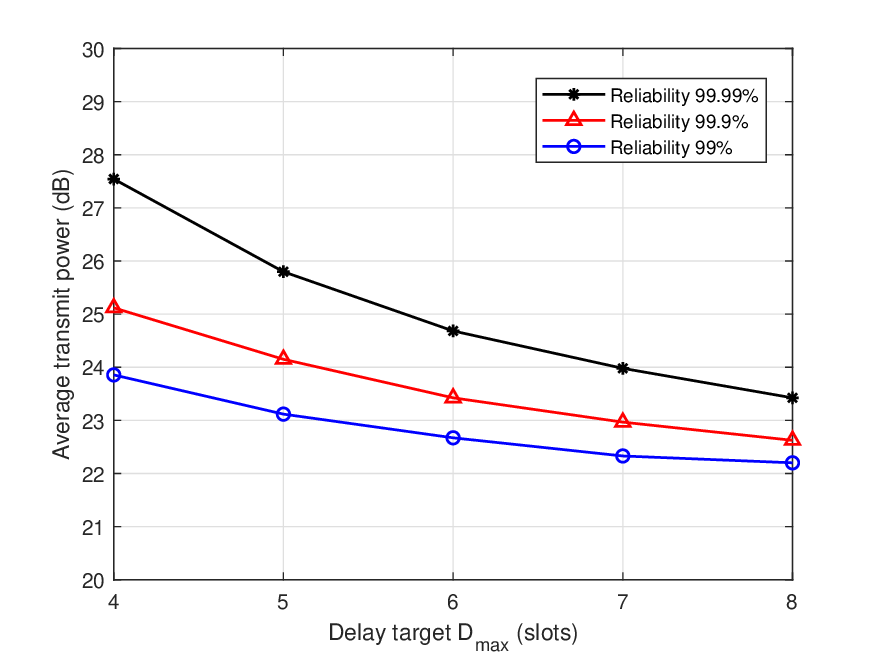}
    \caption{Impact of reliability and latency.}
    \label{fig:tradeoff}
    \vspace{-0.5cm}
\end{figure}

{ Firstly, we show the learning curves of the DRL algorithm in Fig.~\ref{fig:training} by averaging the accumulated reward and delay violation probability from 10 individual training processes with distinct random environment seeds, and the light color region  is the variation region, where the target \gls{dvp} $\xi=10^{-3}$, (i.e., reliability of 99.9\%) and $D_{\rm max} =5$ slots. We can see that the accumulated reward increases drastically in the first few episodes and the simulated \gls{dvp} also decreases drastically, which indicates that there exists a main learning stage. Then in the following, from the 5th-60th episodes, a minor tuning stage exists where the accumulated reward increases slowly and the \gls{dvp} fluctuates around the \gls{dvp} target. After the 60th episode, the accumulated reward remains almost constant and the \gls{dvp} varies within the target \gls{dvp} threshold which indicates that the policy almost converges. In general, Fig.~\ref{fig:training} implies the  \gls{ppo}-based algorithm is capable of solving the transmission control problem with given channel distribution while strictly meeting the \gls{qos} requirement. }

In Fig.~\ref{fig:tradeoff}, we show the impact of the reliability constraint $1-\xi$ and the delay target $D_{\rm max}$ on the average transmit power. Three levels of reliability are considered, namely, $1-\xi=$ 99\%, 99.9\%, and 99.99\%. The delay target $D_{\rm max}$ varies from 4 to 8 slots. 
% Firstly, we demonstrate that the  \gls{ppo}-based algorithm can guarantee the stringent \gls{qos} requirement under variational reliability and latency constraints. Secondly, we can conclude 
It is seen that with higher reliability $1-\xi$ and more stringent delay target $D_{\rm max}$, much more average transmit power is required. This observation coincides with the intuition that, the required transmit power increases more rapidly as the \gls{qos} requirements become more stringent. 

\subsection{Performance Comparison in Synthetic Target Area}

\begin{table}
\begin{center}
\caption{Environment parameter}
\begin{tabular}{|c | c|}
\hline
Simulation Parameters  & Value \\
\hline
Target area  $\mathcal Z$  & $[-20,20]\times [-20,20]$ m$^2$ \\
BS location   & (0,100)~m \\
BS height   & 10 m \\
Device height   & 1.5 m \\
Grid spacing  & 2 m \\
Transmit power for $ \gamma$ acquisition  & 0 dBm \\
Number of total locations  $L$ & 441 \\
Number of known locations  $M$ & 110 \\
Number of channel gain samples  $U$ & $10^3$ \\
Channel model   & 3GPP\_3D\_UMi\_NLOS\\ %QuaDRiGa model
Number of paths    & 12 \\
Central frequency  $f_c$ & 2.6 GHz \\
Bandwidth $B$ & 200 KHz \\
Noise power  $BN_0$ & -115 dBm \\
Stochastic arrival distribution  & Poisson \\
Average arrival rate  $\lambda$  & 800 bits/slot \\
Reliability  1 - $\xi$ & 99.9\% \\
Latency  $D_{\rm max}$ & 5 slots \\
\hline
\end{tabular}
\label{envpara}
\end{center}
\vspace{-0.4cm}
\end{table}

\begin{figure}
% \vspace{-0.2cm}
\centering
\subfigbottomskip=-2pt %两行子图之间的行间距
\subfigcapskip=-5pt %设置子图与子标题之间的距离
\subfigure[Average channel gain $ \gamma$ of the target area averaged over $10^{5}$ channel gain samples ]
{\includegraphics[width=0.82\linewidth]{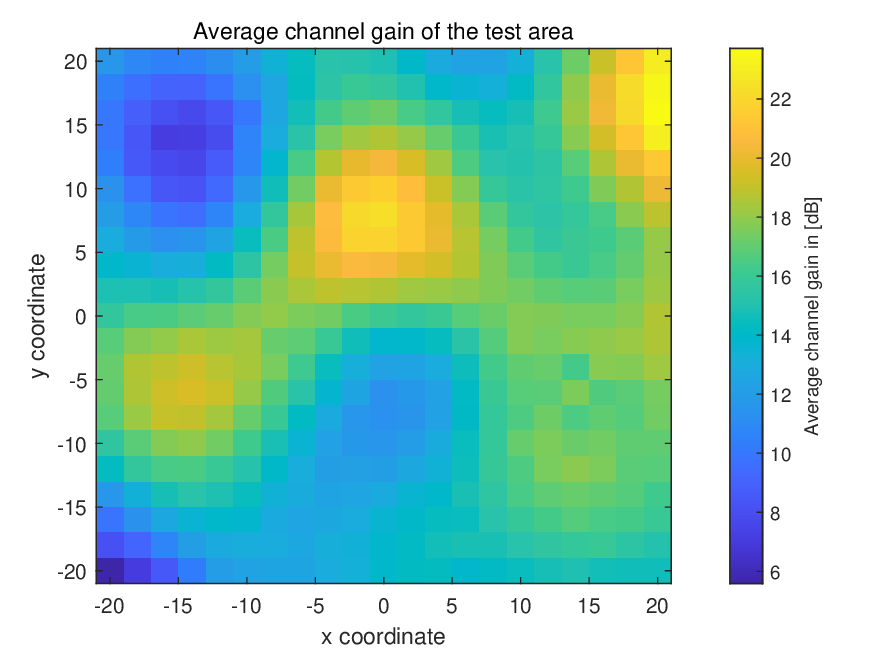}\label{fig:Average-channel-gain}} \\
\subfigure[$\varepsilon$-quantile of the channel gain computed from $10^3$  channel gain samples  ]
{\includegraphics[width=0.82\linewidth]{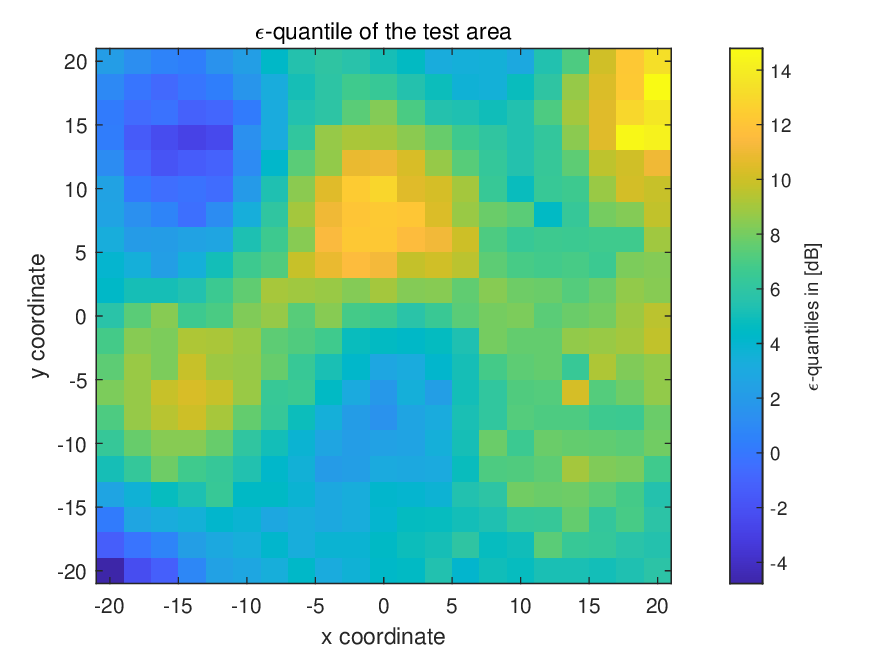}\label{fig:quantile}} \\
\subfigure[$\varepsilon$-quantile predicted  through Gaussian Process from given 110 sampled locations (black dots), each location with $10^3$ channel gain samples.]
{\includegraphics[width=0.82\linewidth]{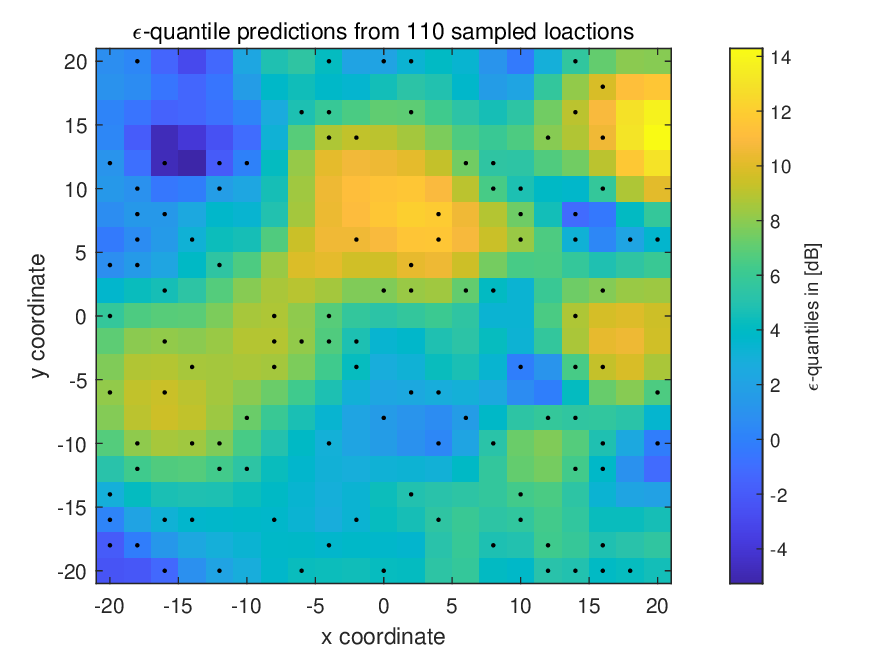} \label{fig:quantile-GP}}
\caption{Average channel gain and the $\varepsilon$-quantile comparison of the target area, where $\varepsilon=0.1$ }
\vspace{-0.5cm}
\end{figure}

% \begin{figure*}
%     \centering
% 	\subfigbottomskip=-1pt %两行子图之间的行间距
% 	\subfigcapskip=0pt %设置子图与子标题之间的距离
% 	\subfigure[Power performance of benchmark 1]{
% 	\includegraphics[width=0.31\linewidth]{Figure/power-baseline1.eps}  \label{fig:benchmark1}   }  
% 	\subfigure[Power performance of benchmark 2]{
% 	\includegraphics[width=0.31\linewidth]{Figure/power-baseline2.eps}  \label{fig:benchmark2}  } 
% 	\subfigure[Power performance of the power scaling solution]{
% 	\includegraphics[width=0.31\linewidth]{Figure/Power-compensation-performance.eps}   \label{fig:scaling}  }   
% 	%\quad
%      \caption{Power performance comparison with a target \gls{dvp} $\xi=10^{-3}$ and $D_{\rm max}=5$ slots, the red dots stand for the $K=8$ base locations with trained policy networks}
%     \label{fig:power comparison}
%     \vspace{-0.4cm}
% \end{figure*}

\begin{figure}
% \vspace{-0.7cm}
    \centering
    \includegraphics[scale=0.58]{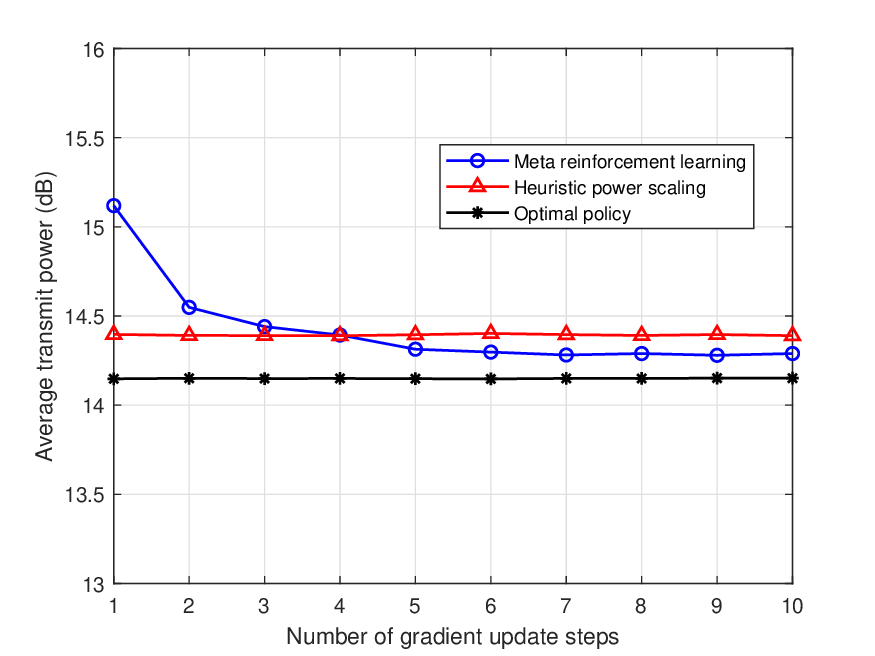}
    \caption{Adaptability convergence performance of the meta-reinforcement learning}
    \label{fig:meta-convergence}
    \vspace{-0.4cm}
\end{figure}

\begin{figure*}[h]
    \centering
	\subfigbottomskip=-1pt %两行子图之间的行间距
	\subfigcapskip=0pt %设置子图与子标题之间的距离
	\subfigure[Power performance of benchmark 1]{
	\includegraphics[width=0.45\linewidth]{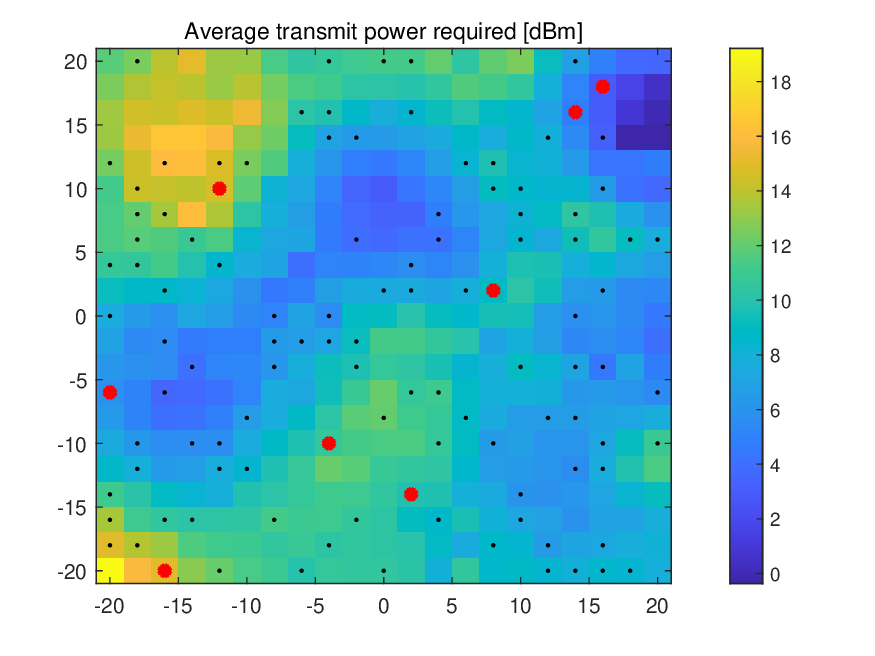}  \label{fig:benchmark1}   }  
	\subfigure[Power performance of benchmark 2]{
	\includegraphics[width=0.45\linewidth]{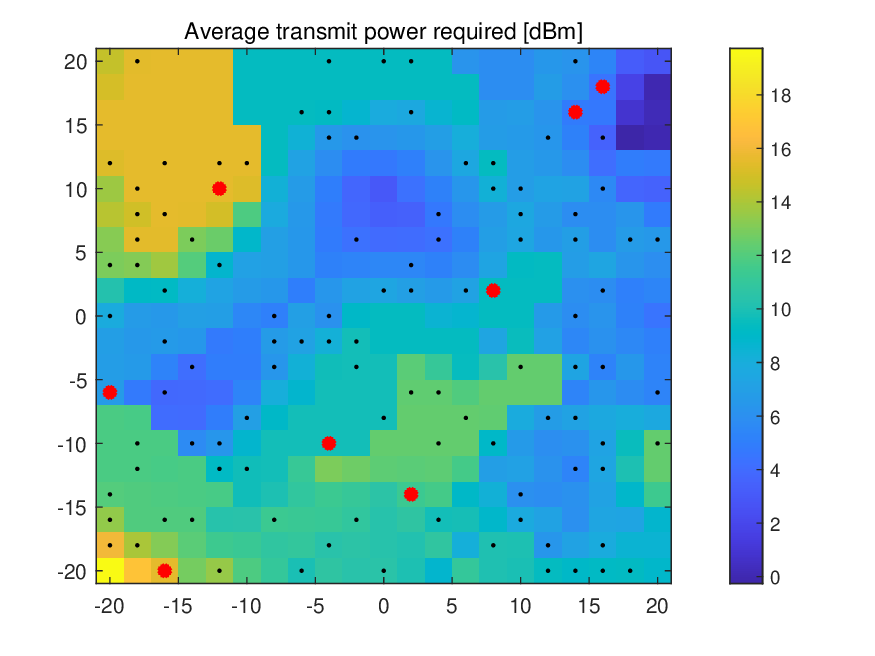}  \label{fig:benchmark2}  } 
	  \\
	\subfigure[Power performance of the power scaling solution]{
	\includegraphics[width=0.45\linewidth]{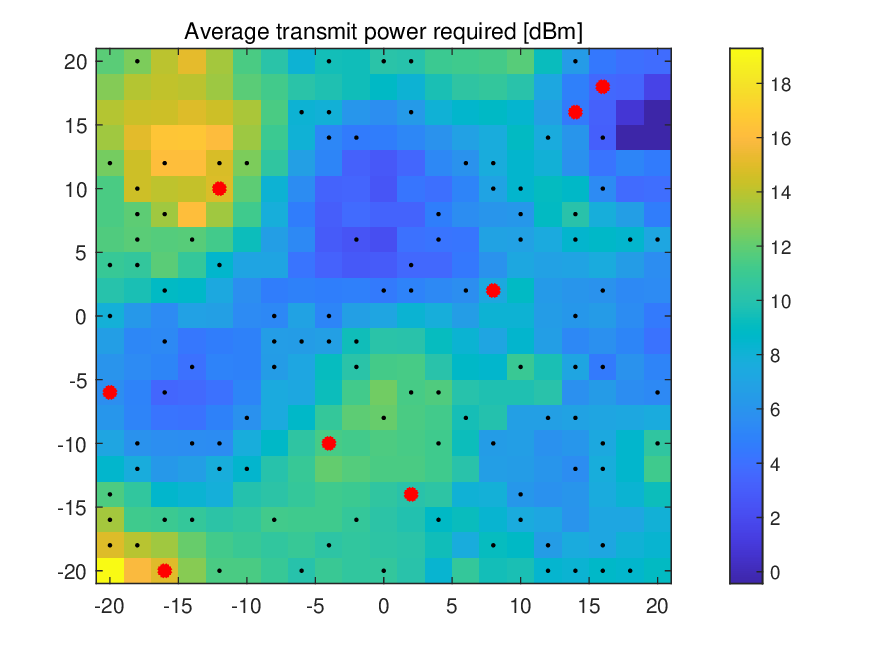}   \label{fig:scaling}  }   
	%\quad
	\subfigure[Power performance of the Meta-DRL solution]{
\includegraphics[width=0.45\linewidth]{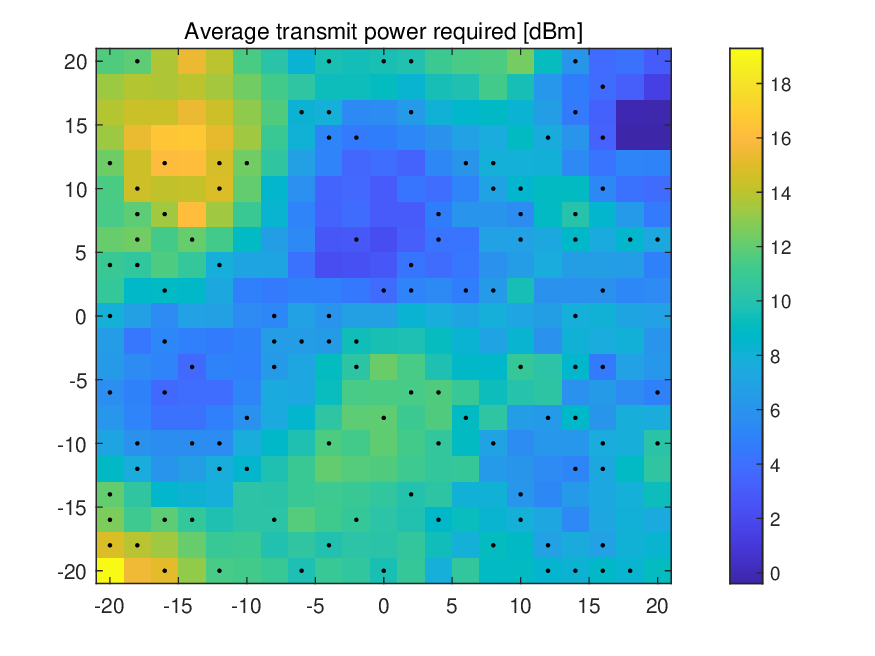}   \label{fig:meta} }    
     \caption{Power performance comparison with a target \gls{dvp} $\xi=10^{-3}$ and $D_{\rm max}=5$ slots, the red dots stand for the $K=8$ base locations with trained policy networks}
    \label{fig:power comparison}
    \vspace{-0.2cm}
\end{figure*}

In this subsection, we will focus on the formulated problem in (\ref{P2}) with a synthetic environment model generated by QuaDRiGa, a well-known channel model generator, in Matlab. 
The details of the synthetic environment are summarized in Table~\ref{envpara}. Specifically, we consider a 21 x 21 grid area, with a total 441 grids. Among these grids, 110 grids with limited channel gain samples are known at the \gls{bs}. At each grid, we generated $10^{5}$ channel gain samples. Among them, $10^3$ channel gain samples are randomly selected as the known channel gain samples at the \gls{bs} and will be used for offline \gls{ckm} establishment and meta-DRL training. {The overall $10^5$ channel gain samples are used for numerical validation as the virtual environment.}

In the following, we first compare the predictive quantile from the \gls{ckm} and the real quantile of this scenario. Then we show the convergence of the meta-reinforcement learning algorithm. Finally, we evaluate the power performance and \gls{dvp} performance of the power scaling approach in conjunction with the \gls{drl} algorithm and the meta-DRL algorithm. {Furthermore, the parameters for the base policies training in the proposed power scaling scheme are the same as in subsection~\ref{Simu-a}.} The only difference is that the peak power cannot be obtained directly but can be calculated as $p_{\rm max}=-\frac{2^{\lambda/n}-1}{\mathbb E\{ \gamma| \mbf x\}  \log[1 - (\xi^{D_{\rm max}^{-1}}/D_{\rm max})] }$. In addition, the number of clusters $K$ is set to be 8 empirically.
The network parameters of the meta-DRL algorithm are also the same as the ideal Rayleigh fading channel in subsection~\ref{Simu-a}. The number of tasks $N$ is set as 20. In addition, the meta-learning rate $\beta$ is set as $10^{-3}$.

\subsubsection{Validation of the predictive \gls{ckm}}

In Fig.~\ref{fig:Average-channel-gain}, we plot the average channel gain of the target area, obtained by averaging the $10^5$ channel gain samples of each location. Note that there are three power peaks located around (20, 16), (0,10), and (-15, -5) and two power valleys around (-20,-20) and (-13, 13) for the particular realization of our simulated environment. 
In Fig.~\ref{fig:quantile}, we plot the 0.1-quantiles of the known channel gain samples. We can see that the overall shape is similar to the average channel gain with a rough grid and different average values. Fig.~\ref{fig:quantile-GP} depicts the predicted 0.1-quantile \gls{ckm} through 110 locations marked as dots on the map. We can see that this map is also similar to Fig.~\ref{fig:quantile} except for those locations far away from the observed locations in $\mathcal D$. It is also seen that the predicted 0.1-quantile \gls{ckm} is quite similar to the real one due to the space consistency through continuous Gaussian process interpolation.  

\begin{figure*}[h]
    \centering
	\subfigbottomskip=-1pt %两行子图之间的行间距
	\subfigcapskip=0pt %设置子图与子标题之间的距离
	\subfigure[\gls{dvp} performance of benchmark 1]{
	\includegraphics[width=0.45\linewidth]{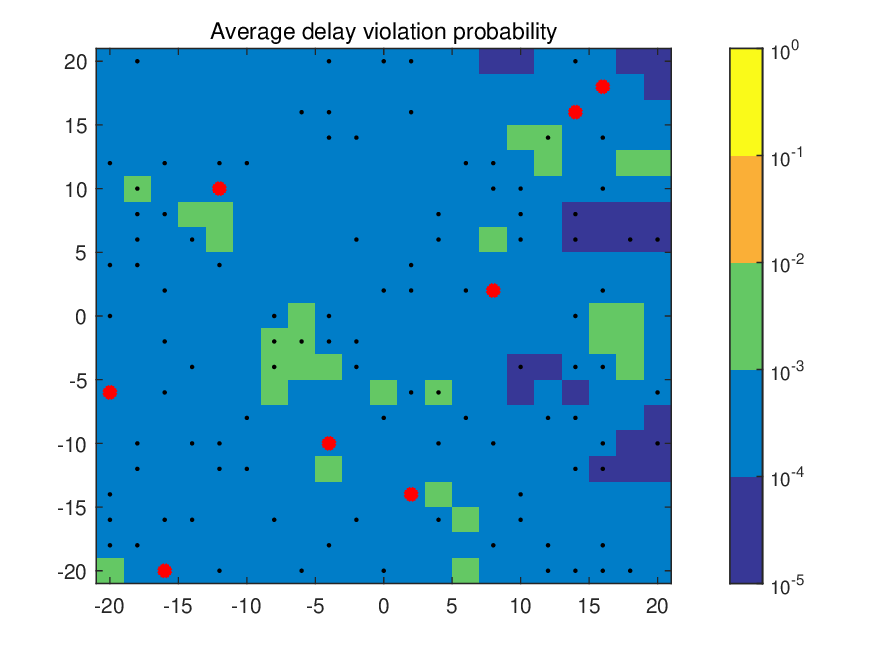}  \label{fig:benchmark1-dvp}   }  
	\subfigure[\gls{dvp} performance of benchmark 2]{
	\includegraphics[width=0.45\linewidth]{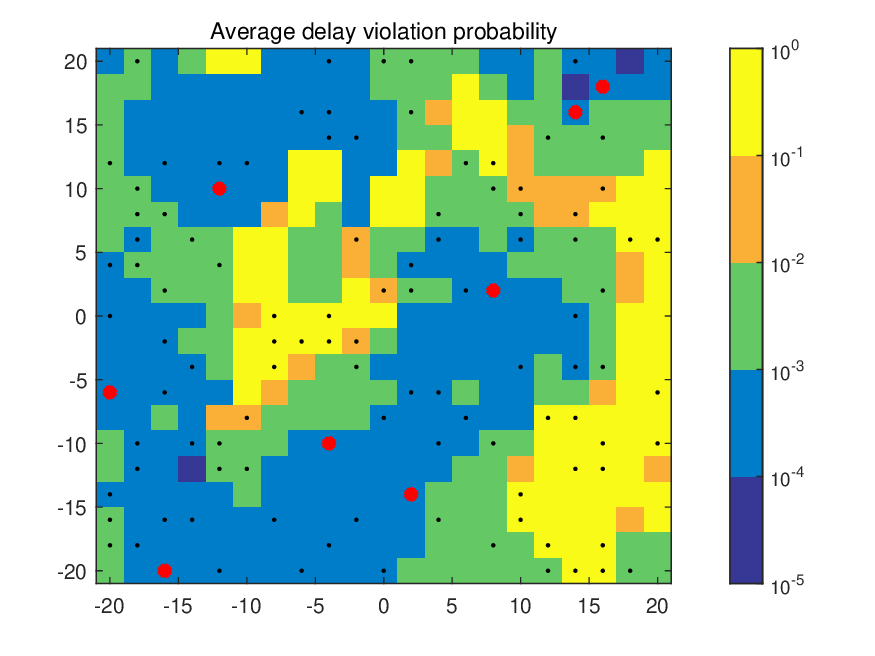}  \label{fig:benchmark2-dvp}  } 
	  \\
	\subfigure[\gls{dvp} performance of the power scaling solution]{
	\includegraphics[width=0.45\linewidth]{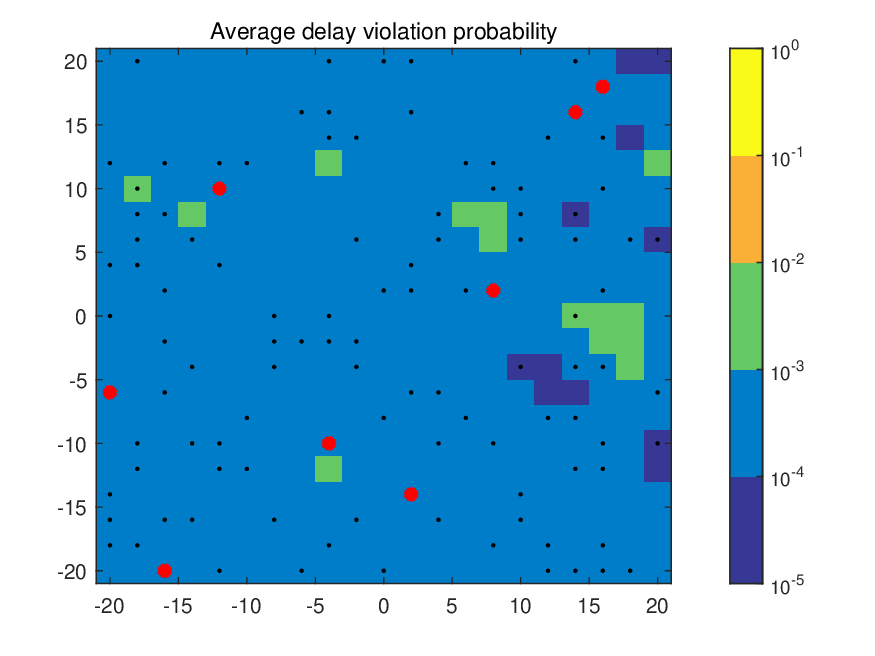}   \label{fig:scaling-dvp}  }   
	%\quad
	\subfigure[\gls{dvp} performance of the Meta-DRL solution]{
	\includegraphics[width=0.45\linewidth]{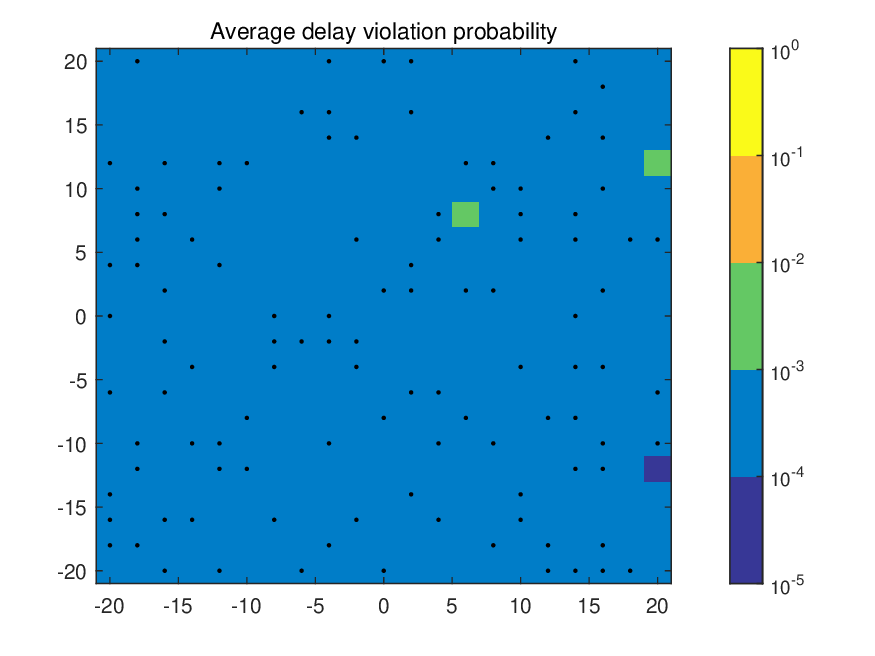}   \label{fig:meta-dvp} }    
     \caption{\gls{dvp} performance comparison with a target \gls{dvp} $\xi=10^{-3}$ and $D_{\rm max}=5$ slots, the red dots stand for the $K=8$ base locations with trained policy networks}
    \label{fig:dvp comparison}
    \vspace{-0.2cm}
\end{figure*}

\subsubsection{Convergence validation of the meta-\gls{drl} algorithm}
To show the adaptability of the meta-reinforcement learning algorithm, we first show the convergence performance of the meta-policy in a Rayleigh fading scenario with average channel gain $\mathbb E \{ \gamma\} = 10$dB as the gradient update steps increase in Fig.~\ref{fig:meta-convergence}. Here the gradient is updated every $T_{\text {ap}}=1000$ time slots as (\ref{PPOP}) in the adaptation stage.
Simultaneously, we also show the performance of the power scaling solution by selecting an appropriate base location. {The average transmit power of the meta-reinforcement learning algorithm initialization without parameter update conducted from the interaction with the environment does not outperform the proposed power scaling scheme.} As the number of gradient updates increases, the average transmit power decreases drastically and then gradually converges within 10 gradient update steps and outperforms the proposed power scaling scheme. The performance of the power scaling solution and the optimal policy are almost constant since these two schemes can not learn from the new environment.

\subsubsection{Validation of the proposed solutions}
In order to evaluate the performance of the two solutions, we introduce two benchmark schemes for comparison. The details of these solutions are summarized as follows:
\begin{itemize}
\item  {Meta-\gls{drl} solution:} This solution shows the adaptation result of the meta-\gls{drl} solution as in Algorithm~\ref{Alg3} after 10 gradient update steps adaptation in the new locations. 
\item  {Power scaling solution:} This solution shows the result of power scaling scheme through the improved $K$-means clustering. 
\item  {Benchmark 1:} This solution also shows the result of the power scaling scheme, but rather than the improved $K$-means clustering method, it simply uses the policy from the nearest base location on the map for power scaling. %The clusters of the $K=8$ base locations are different for this method. %Power scaling solution through nearby base location:
\item  {Benchmark 2:} This solution directly utilizes the nearest base location's policy for the current location on the map without any adaptation. %Nearby base-policy location:
\end{itemize}

Fig.~\ref{fig:power comparison} depicts the transmit power performance of these four schemes.  Here we randomly choose one channel gain sample from the $10^5$ generated channel gain samples to interact with the agent in each time slot and calculate the corresponding average transmit power and the overall \gls{dvp} over $10^5$ slots for the locations within the target area. It is worth mentioning that we still use black dots to stand for the sampled locations and red dots to stand for base locations on the map in Fig.~\ref{fig:power comparison} and Fig.~\ref{fig:dvp comparison}.
It is observed that the transmit power maps of these four schemes are roughly complementary to the average channel gain map in Fig.~\ref{fig:Average-channel-gain}.  The power performance of the two solutions and benchmark 1 are quite similar. While the performance of benchmark 2 is worse than the above 3 solutions. 

In Fig~\ref{fig:dvp comparison}, we show the \gls{dvp} performance maps of the four schemes. Specifically, we use different colors for different orders of magnitude of the \gls{dvp}. 
For benchmark 1, with the help of the power scaling scheme, the \gls{dvp} constraint is improved greatly, and only a small amount of locations' \gls{dvp} constraints are violated by one order of magnitude as depicted in Fig.~\ref{fig:benchmark1-dvp}. 
For benchmark 2, \gls{dvp} constraints of more than half of the locations within the target area are not met as depicted in Fig.~\ref{fig:benchmark2-dvp}. The \gls{dvp} constraints are violated by a few orders of magnitude. More specifically,  most of these locations under \gls{dvp} constraints are around the base-policy locations, which indicates the spatial correlation of the channel distribution. 
For the proposed power scaling scheme in Fig.~\ref{fig:scaling-dvp}, the \gls{dvp} performance is better than benchmark 1 due to the improved $K$-means clustering. 
Finally, the \gls{dvp} performance of the meta \gls{drl} based algorithm outperforms the \gls{drl}-based power scaling scheme, and only two locations' \gls{dvp} constraints are violated as shown in Fig.~\ref{fig:meta-dvp}. 
 
As a result, we calculate the average transmit power and corresponding availability across the target area through different solutions in Table~\ref{finalcomp}. The average transmit power is calculated by averaging the long-term average transmit power across the target area through different solutions. The availability is calculated by the probability that the QoS (i.e., reliability and latency) of users can be satisfied in a wireless network. In the space domain, it is the ratio of the covered area, within which the QoS can be satisfied, to the total service area. Furthermore, to be more persuasive, we also show the 95\% confidence interval of the availability by normal approximation. This confidence interval is an interval estimate of a success probability calculated from the outcome of the overall $L=441$ binomial experiments, i.e., \gls{dvp} performance evaluation. 
In general, we can conclude that by directly applying the $K$=8 trained networks into the considered area without transmit power scaling, the overall availability only achieves 44.22\%. When the \gls{ckm} is established, the transmit power scaling from most nearby base-policy can significantly enhance the availability to 93.42\%. In the meanwhile, the average transmit power is reduced by about 14\%. When both the improved $K$-means clustering and transmit power scaling are employed, both the transmit power and availability are slightly improved, which implies the effectiveness of the $K$-means clustering. Finally, we can observe that the performance of the meta-\gls{drl} method can achieve the best power performance and the highest availability, which shows the effectiveness of the meta-\gls{drl} algorithm. 

\begin{table*}
\begin{center}
\caption{Performance comparison}
\begin{tabular}{|c|c|c|c|c|}
\hline
Solutions & Meta-DRL & Power scaling & Benchmark1 & Benchmark2 \\
\hline
Average transmit power & 8.978 mW (9.532 dBm) & 9.034 mW (9.559 dBm) & 9.378 mW (9.721 dBm) & 10.689 mW (10.289 dBm) \\
\hline
 Availability observation& 99.55\% & 96.83\% & 92.97\%  & 44.22\%\\
\hline
95\% confidence interval of the availability &  98.93\% $\sim$ 100\% & 95.19\% $\sim$ 98.47\% & 90.62\% $\sim$ 95.38\% & 39.58\% $\sim$ 48.86\%\\
\hline
\end{tabular}
\label{finalcomp}
\end{center}
\vspace{-0.4cm}
\end{table*}

\section{Conclusion}\label{sec7}
We have investigated the transmission control adaptation problem for mission-critical \gls{iot} systems with \gls{urllc} in a target area exploiting the channel gain samples of a few locations known at the \gls{bs}. 
We have first formulated a transmission control problem for a specific location with known channel gain samples while guaranteeing the stringent \gls{qos} requirement of \gls{urllc} by dynamically determining the transmit power and coding rate. This was followed by a policy adaptation problem for the users across the target area with various channel statistics under the same \gls{qos} requirement. We have first utilized a \gls{ppo}-based algorithm to solve the transmission control problem for a specific location. This has been supplemented by a power scaling scheme based on the above \gls{ppo}-based algorithm via \gls{ckm} for the policy adaptation problem without re-training. 
% and a meta-\gls{drl} based algorithm are opposed to solve the policy adaptation problem. 
% To solve the policy adaptation problem across the target area, we proposed a power scaling scheme based on \gls{drl}-based algorithm and a meta-\gls{drl} based algorithm. The \gls{drl}-based heuristic algorithm first establishes a \gls{ckm}  of the target area and then divides all the locations into $K$ clusters through improved $K$-means clustering to further improve the power efficiency. 
% To further improve the power efficiency, we divide all the locations into $K$ clusters through improved $K$-means clustering to mitigate the distribution mismatch by selecting more correlated $\mathscr E$-quantiles into the same cluster. 
%through Gaussian process which can predict the corresponding $\mathscr E$-quantiles of all the locations within the target area.  
% Then we conduct power scaling based on the established  \gls{ckm} and corresponding clusters without the need for additional training time. 
Our second proposal is \gls{maml}-based meta-reinforcement learning algorithm based on the \gls{ppo}-based algorithm, which can effectively train a meta policy. This meta policy can adapt quickly to new environments within a few gradient updates after interacting with the realistic environment.
The numerical results have validated the effectiveness of the PPO-based algorithm for the transmission control problem under various \gls{qos} requirements. It has been shown that the two proposals can solve the transmission control adaptation problem with high \gls{qos} availability compared to the benchmark scheme without power scaling. {The immediate future work is a generalization of the proposed methods to the base stations and UEs with multiple antennas.}
{Furthermore, future work can also take the following two practical impacts into account: The first is the user mobility within the target area. The second is maintaining the dynamic  \gls{ckm}, since the realistic channel models usually comprise of non-static propagation environments. 
}

%and the corresponding \gls{dvp} performance of these two solutions is significantly better than the benchmark scheme without power scaling, indicating the effectiveness of the two proposed solutions. 

\bibliographystyle{IEEEtran}
\bibliography{reference}
\end{document}